\documentclass[10pt,final,conference,a4paper,twoside]{IEEEtran}

\usepackage{times}
\usepackage{amsmath,amssymb,amsbsy,stmaryrd,shuffle}
\usepackage{galois}
\usepackage[normalem]{ulem}
\usepackage{todonotes}
\usepackage{pgf}
\usepackage{tikz}
\usetikzlibrary{matrix,arrows,automata}

\usepackage[pdftex,%
 	colorlinks,%
	citecolor=black,
 	pdfauthor={Javier Esparza, Pierre Ganty, Rupak Majumdar},%
 	hypertexnames=false,%
	pdftitle={A Perfect Model for Bounded Verification}]{hyperref}

\newif
  \iflong
  \longtrue
\newif
  \ifshort
	\shortfalse

\def\rtape{\ensuremath{\mathsf{Rgt}}}
\def\ltape{\ensuremath{\mathsf{Lft}}}

\def\tapes{\mathfrak{T}}
\def\tape{\bar{t}}

\def\set#1{{\{ #1 \}}}

\def\denote#1{{\llbracket #1 \rrbracket}}
\def\tuple#1{{\langle #1 \rangle}}
\def\nats{{\mathbb{N}}}
\def\integers{{\mathbb{Z}}}

\def\pzero{\mathbf{0}}
\def\punit{\mathbf{e}}

\def\post{\ensuremath{\mathrm{post}}}
\def\op{\mathit{op}}
\def\cfsm{\mathsf{CFSM}}
\def\trans{{\ensuremath{\mathcal{T}}}}

\def\readh#1{\left[ {#1}\right\rangle}
\def\enqueue#1{!{#1}}
\def\dequeue#1{?{#1}}
\def\channel{\mathcal{C}}

\def\pat{\bar{\boldsymbol{w}}}
\def\pata{\bar{\boldsymbol{a}}}
\def\patb{\bar{\boldsymbol{b}}}

\def\rmark{\$}

\def\trpda#1{\ensuremath{\overset{{#1}}{\hookrightarrow}}}

\def\prod{\mathcal{P}}

\def\parikh{\ensuremath{{\mathsf{Parikh}}}}
\def\proj{{\pi}}

\def\pzero{\mathbf{0}}
\def\punit{\mathbf{e}}

\def\readh#1{\left[ {#1}\right\rangle}

\def\kcm{\mathsf{CM}}
\def\config{{\mathbf{c}}}

\def\inc{\mathsf{inc}}
\def\dec{\mathsf{dec}}
\def\zerotest{\mathsf{zerotest}}

{\bfseries\upshape}{\itshape}
\newtheorem{@protheo}{Theorem}
\newenvironment{theorem}[1]{\begin{@protheo}{\rm \bf #1}\it}{\end{@protheo}}
\newtheorem{remark}{Remark}{\bfseries\upshape}{\rm}
\newtheorem{definition}{Definition}{\bfseries\upshape}{\itshape}
\newtheorem{proposition}{Proposition}{\bfseries\upshape}{\itshape}
\newtheorem{lemma}{Lemma}{\bfseries\upshape}{\itshape}
{\bfseries\upshape}{\itshape}

\newcommand{\bwt}{\!\bowtie\!}
\newcommand{\ish}{\underline{\shuffle}}
\newcommand{\oSigma}{\overline{\Sigma}}

\begin{document}
\title{A Perfect Model for Bounded Verification}

\author{\IEEEauthorblockN{Javier Esparza}
\IEEEauthorblockA{TUM}
\and
\IEEEauthorblockN{Pierre Ganty}
\IEEEauthorblockA{IMDEA Software Institute}
\and
\IEEEauthorblockN{Rupak Majumdar}
\IEEEauthorblockA{MPI-SWS}}

\maketitle

\begin{abstract}
A class of languages $\mathcal{C}$ is {\em perfect} if it is closed under
Boolean operations and the emptiness problem is decidable.  Perfect language
classes are the basis for the automata-theoretic approach to model checking: a
system is correct if the language generated by the system is disjoint from the
language of bad traces.  Regular languages are perfect, but because the
disjointness problem for context-free languages is undecidable, no class
containing them can be perfect.

In practice, verification problems for language classes that are not perfect
are often under-approximated by checking if the property holds for all
behaviors of the system belonging to a fixed subset.  A general way to specify
a subset of behaviors is by using bounded languages (languages of the form
$w_1^*\ldots w_k^*$ for fixed words $w_1,\ldots,w_k$).  A class of languages
$\mathcal{C}$ is {\em perfect modulo bounded languages} if it is closed under
Boolean operations relative to every bounded language, and if the emptiness
problem is decidable relative to every bounded language.

We consider finding perfect classes of languages modulo bounded languages.  We
show that the class of languages accepted by multi-head pushdown automata are
perfect modulo bounded languages, and characterize the complexities of decision
problems.  We also show that bounded languages form a maximal class for which
perfection is obtained.  We show that computations of several known models of
systems, such as recursive multi-threaded programs, recursive counter machines,
and communicating finite-state machines can be encoded as multi-head pushdown
automata, giving uniform and optimal underapproximation algorithms modulo
bounded languages.
\end{abstract}

\section{Introduction}

The automata-theoretic approach to model checking linear-time
properties formalizes the verification problem as a language-theoretic problem about
two automata: the {\em system automaton}, which recognizes the set of executions of the system,
and the {\em property automaton}, which recognizes either the sequences of actions
satisfying the property ({\em positive} specification), or those violating it ({\em negative} specification). Given a system automaton $S$ and a property automaton $P$, verification of positive and negative
specifications reduces to checking $L(S) \subseteq L(P)$ (inclusion problem), or to checking $L(S) \cap L(P) = \emptyset$ (disjointness problem) , respectively.

Language classes effectively closed under boolean operations
and with a decidable emptiness problem are particularly interesting for the automata-theoretic approach.
For such classes not only the inclusion and disjointness problems are decidable, they also have many further
advantages. For example, in these classes systems are closed under parallel composition by rendez-vous, properties are closed under boolean operations, and systems can be seen as properties, or vice versa,
with many useful consequences for compositional and assume-guarantee verification
techniques. For all these reasons, we call these classes {\em perfect}.

The regular languages are perfect but, since  because the disjointness problem for
the context-free languages (CFL) is undecidable (see \cite{HU79}), no class containing CFL can be perfect. This ``context-free barrier'' restricts the search for perfect
classes to those properly contained in CFL or incomparable with them, and both possibilities have been investigated. In a seminal paper \cite{DBLP:journals/jacm/AlurM09},
Alur and Madhusudan proved that the visibly pushdown languages----a subclass of CFL---are perfect,
a result that lead to a very successful theory and efficient algorithms (see e.g.\cite{DBLP:conf/fsttcs/LodingMS04,DBLP:journals/jacm/AlurM09}). Later La Torre, Madhusudan, and Parlato discovered a perfect class incomparable with CFL:  the languages recognized by multi-stack visibly pushdown
automata whose computations can be split into a fixed number of stages during which at most one stack is popped \cite{DBLP:conf/lics/TorreMP07}.

The ``context-free barrier'' continues to be a serious obstacle in many applications, in particular in the verification
of concurrent systems.  For this reason, many tools only check a subset of the executions of the system.
Intuitively, they direct a {\em spotlight} to a region of the possible executions, and check whether the
executions {\em under the spotlight} satisfy the property. The spotlight is controlled by the user,
who can freely move it around to check different regions, and conventional verification corresponds to a
spotlight that illuminates all the space of possible executions. In particular, the ``spotlight principle''
is applied by bounded model-checkers, which unroll program loops and recursion up to a
fixed depth (often after taking the product of the program with an automaton
for the property to be checked), leaving a system
whose executions have a fixed bounded length
(see e.g. \cite{DBLP:journals/fmsd/ClarkeBRZ01,DBLP:conf/dac/ClarkeKY03}).
It is also used by {\em context-bounded} checkers for multi-threaded programs \cite{QR,Shaz08,DBLP:journals/corr/abs-1111-1011}, which only examine
executions containing at most a fixed number of context-switches (communication events between threads).
Context-bounded checkers break the context-free barrier, but at the price of only exploring finite action sequences.\footnote{
	More precisely, in automata-theoretic terms context-bounded checkers explore runs of $S$ of arbitrary length, but containing only a fixed number of
	non-$\varepsilon$ transitions.}
Recently, building on ideas by Kahlon \cite{patent:WO/2009/094439}
on bounded languages \cite{ginsburg},
context-bounded checking has been extended to {\em bounded verification} \cite{ge11}\footnote{In \cite{ge11} bounded verification was called
pattern-based verification, but, since pattern is a rather generic term, we
opt for bounded verification here.},
which checks whether executions of the system of the form $w_1^* \ldots w_n^*$ for some
finite words $w_1, \ldots, w_n$ satisfy a property.

In automata-theoretic terms, the spotlight principle corresponds to {\em verification modulo a language}.
The inclusion check $L(S) \subseteq L(P)$ and the disjointness check $L(S) \cap L(P) = \emptyset$ are replaced by
checks $L_M(S)  \subseteq L_M(P)$ and $L_M(S)  \cap L_M(P)  = \emptyset$, respectively, where $L_M$ denotes $L \cap M$.
Context-bounded checking corresponds to verification modulo the language of all words up to fixed length, and bounded verification to verification modulo a bounded expression.

Verification modulo a language $M$ allows to break the context-free barrier, which raises the question of
identifying perfect classes {\em modulo language classes}.
Given a boolean operation ${\it Op}(L_1, \ldots, L_n)$ on languages,
let us define the same operation modulo a language $M$ by ${\it Op}_M(L_1, \ldots, L_n) = {\it Op}(L_1\cap M, \ldots, L_n \cap M)$,
and, similarly,  let us say that an automaton $A$ is empty modulo $M$  if $L(A) \cap M = \emptyset$.
Let $\mathcal{L}$ and $\mathcal{C}$ be classes of languages.
We call $\mathcal{L}$ {\em perfect modulo $\mathcal{C}$} if it is closed under Boolean operations modulo any $M\in\mathcal{C}$,
and has a decidable emptiness problem modulo any $M\in \mathcal{C}$.
It is easy to see that the recursive languages are perfect modulo the finite languages.
But for bounded expressions the question becomes harder.
The disjointness problem modulo a bounded expression is decidable for CFL \cite{ginsburg},
which hints at a perfect class modulo bounded expressions containing CFL.
However, CFL itself is not perfect modulo bounded expressions, because it is not closed under intersection:
there is no CFL $L$ such that $\set{a^nb^nc^*\mid n\geq 0} \cap \set{a^*b^nc^n\mid n\geq 0}\cap a^*b^*c^* = L\cap a^*b^*c^*$.

In this paper we present the first perfect class modulo bounded expressions: the
languages recognized by {\em multihead pushdown automata} (MHPDA).
This result is very satisfactory, because the class has a simple and purely syntactic definition,
and as we demonstrate, is expressive enough to capture many well-known models.
We also characterize the complexity of the Booleans operations and the emptiness check modulo bounded expressions:
we show that the emptiness check is coNEXPTIME-complete, union and intersection are polynomial,
and complementation is at most triply exponential.
Surprisingly, the emptiness problem is coNP-complete (and complementation doubly
exponential)
for the subclass of {\em letter-bounded} expressions, in which
each string $w_1,\ldots, w_n$ is a single letter.
We also show that bounded expressions are a maximal class of regular languages for which perfection
can be attained for MHPDAs, any additional language leads to undecidability of emptiness.

In the second part of the paper, we show that central automata models of
software can be encoded into MHPDA.  Encoding recursive multithreaded programs
to MHPDA is obvious, since the intersection of CFLs is MHPDA-definable, and we
subsume the results of Esparza and Ganty \cite{ge11}.  Additionally, we supply
encodings for recursive counter machines (\(\kcm\)), the main
automata-theoretic model of procedural programs with integer variables, and for
finite-state machines communicating through unbounded perfect FIFO channels
(\(\cfsm\)), the most popular model for the verification of communication
protocols.  While the existence of some encoding is not surprising, since
emptiness problems for \(\kcm\), \(\cfsm\), and MHPDA are all undecidable, our
encodings exhibit only a small polynomial blowup, and, perhaps more
importantly, preserve bounded behaviours. More precisely, using our encodings
we reduce {\em bounded control-state reachability} for \(\kcm\) and
\(\cfsm\)---deciding reachability of a given control state by means of a
computation conforming to a bounded expression---to non emptiness of MHPDA
modulo bounded expression.  As a consequence, we prove that bounded
control-state reachability for both \(\kcm\) and \(\cfsm\) are NP-complete. The
NP-completeness also extends to {\em unrestricted} control-state reachability
for {\em flat} \(\kcm\) and \emph{flat} \(\cfsm\), because by construction
their computations conform to a bounded expression. (See e.g.
\cite{Leroux05flatcounter} and \cite{bouhabtcs} for a study of those models).
More generally, our language-based approach provides a uniform framework for
the verification of models using auxiliary storage like counters, queues or a
mix of both as defined in \cite{BFatva04}.  Incidentally, our framework allows
to uniformly derive optimal complexity upper bounds for models manipulating
counters, queues or both, and shared memory multithreaded programs.

\smallskip\noindent{\em Related work.} Multi-tape and multi-head finite-state
and pushdown machines were extensively studied in the 1960's and 1970's, e.g.
\cite{IbarraPhD67,ib1974,soudb1974}.
The decidability of emptiness for MHPDA modulo bounded languages
was proved by Ibarra in \cite{ib1974}, using previous results going back to his
(hard to find) PhD thesis \cite{IbarraPhD67}. Our proof settles the complexity
of the problem (coNEXPTIME-complete). Additionally, our constructions show
the surprising coNP-completeness result for letter-bounded expressions.
(A similar coNP-completeness result was recently
obtained in \cite{HL11}, but for a different model.)


Reversal bounded counter machine as bounded language acceptors (see e.g. \cite{ibarra78}) and Bounded Parikh
automata \cite{CFM-words11} have the same expressive power as MHPDA modulo
bounded expressions (they all recognize the languages of the form $\{w_1^{k_1}
\ldots w_n^{k_n} \mid (k_1, \ldots, k_n) \in S\}$ for some semilinear set $S$).
These three characterizations of the same class complement each other. While
MHPDAs have the modelling advantage of allowing to directly encode recursive
procedures, queues and counters, reversal bounded counter machine (and by extension
flat counter machine) have very good
algorithmic methods and tool support (see e.g.
\cite{HL11}\cite{DBLP:journals/sttt/BardinFLP08}).  Our results allow to apply these
algorithms and tools to a larger range of problems.

\section{Preliminaries}

\noindent\emph{Language theory.}
An alphabet \(\Sigma\) is a finite and non-empty set of letters.  We use
$\Sigma^*$ for the set of finite words over $\Sigma$, $\varepsilon$ for the
empty word.

We assume the reader is familiar with the basics of language theory, such as
regular languages, context-free languages (CFL), context-sensitive languages
(CSL),  and the formalisms to describe them: nondeterministic finite automata
(NFA), context-free grammars (CFG), pushdown automata (PDA), etc. (see, e.g.,
\cite{HU79}).

Let us mention that for NFAs, CFGs and PDAs the size of their encoding (denoted
using \(|\cdot|\)) is the number of bits required to represent them.

\smallskip
\noindent\emph{Parikh images.} For \(k\in\nats\), we write \(\integers^k\) and \(\nats^k\) for
the sets of (\(k\)-dim) vectors of integers and naturals,
\(\pzero\) for \((0,\dots,0)\), and \(\punit_i\) for the vector \((z_1,\dots,z_k)\in\nats^k\)
such that \(z_j = 1\) if \(j = i\) and \(z_j = 0\) otherwise.
\emph{Addition} and \emph{equality} on \(k\)-dim vectors are defined pointwise.

Given a fixed linear order \(\Sigma=\set{a_1,\ldots,a_n}\),
the \emph{Parikh image} of \(a_i\in\Sigma\),
written \(\parikh^{\Sigma}(a_i)\), is the vector \(\punit_i\).
The Parikh image is extended to words by defining
\(\parikh^{\Sigma}(\varepsilon)=\pzero\) and \(\parikh^{\Sigma}(u\cdot
v)=\parikh^{\Sigma}(u)+\parikh^{\Sigma}(v)\), and to languages by letting
\(L\subseteq\Sigma^*\), \(\parikh^{\Sigma}(L)=\set{\parikh^{\Sigma}(w)\mid w\in L}\).
We sometimes omit the superscript \(\Sigma\).

\smallskip
\noindent\emph{Presburger Formulas.} A {\em term} is a constant \(c\in\nats\), a variable \(x\) from a set \(X\) of variables, or an expression of the form \(t_1 + t_2\) or \(t_1 - t_2\),
where \(t_1, t_2\) are terms. A {\em Presburger formula} is an expression of the form \(t \sim 0\),
where \(t\) is a term and \({\sim}\in\set{\leq, <, =, \neq, >, \geq}\), or of the form \(\phi_1\wedge
\phi_2\), \(\phi_1\vee \phi_2\), \(\exists x.\phi\), \(\forall x.\phi\), where
\(\phi\), \(\phi_1\), \(\phi_2\) are Presburger formulas.  Given a Presburger formula \(\phi\) with free
variables \(x_1,\ldots,x_k\) (written $\phi(x_1, \ldots, x_k)$), we denote by
\(\denote{\phi}\) the set \(\set{(n_1,\ldots,n_k)\in\nats^k \mid \phi(n_1,\ldots,n_k) \text{ is true}}\),
where $\phi(n_1,\ldots,n_k)$ denotes the formula without free variables obtained by substituting
$n_i$ for $x_i$.
We recall that satisfiability of Presburger formulas is decidable \cite{ginsburg}
and that the Parikh image of a context-free language is Presburger-definable
\cite{Seidl05}.

\smallskip
\noindent\emph{Bounded expressions.\label{subsec:be}}
A \emph{bounded expression} \(\pat\) over \(\Sigma\) is a
regular expression of the form \(w_1^* \dots w_n^*\) such that \(n\geq 1\) and \(w_i\) is a
non-empty word over \(\Sigma\) for each \(i\in[1,n]\)\footnote{For integers \(x\leq x'\), we write \([x,x']\) for the set
\(\set{i\in\integers\mid x\leq i\leq x'}\).}. Abusing notation we sometimes
write \(\pat\) for \(L(\pat)\). The {\em size} of a bounded expression $\pat$ is defined
as $|\pat|=1+\sum_{i=1}^n |w_i|$.
A bounded expression is {\em letter-bounded} if $|w_1| = \ldots = |w_n| = 1$, where
the $w_i$s are not necessarily distinct.

\smallskip
\noindent\emph{Shuffle and indexed shuffle.}
The {\em shuffle} of two words \(x,y\in\Sigma^*\) is the language
\begin{multline*}
x\shuffle y =\{x_1y_1\dots x_ny_n\in\Sigma^* \mid \mbox{each }x_i,y_i \in\Sigma^* \\
\mbox{ and }x=x_1\cdots x_n\, \land y=y_1\cdots y_n\}\enspace .
\end{multline*}
\noindent and the shuffle of two languages $L_1, L_2 \subseteq \Sigma^*$
is the language \(L_1\shuffle L_2 = \textstyle{\bigcup_{x\in L_1, y\in L_2}} x\shuffle y\).
Shuffle is associative, and so we can write $L_1 \shuffle \ldots \shuffle L_k$,
which we often shorten to $\shuffle_{i=1}^k L_i$.

Given \(i>0\) let \( \Sigma \bwt i = \set{\tuple{\sigma,i}\mid \sigma\in\Sigma}\). We
say that $i$ is the {\em index} of $\tuple{\sigma,i}$, and extend indexing to words and
languages in the natural way. For \(w=b_1 \dots b_t\in\Sigma^*\) and \(i > 0\),
\( (w\bwt i) = \tuple{b_1,i}\cdots \tuple{b_t,i}\), and
\(L\bwt i=\set{ w\bwt i\mid w\in L}\). The {\em indexed shuffle} of $L_1, \ldots, L_k$
is the language
$$\ish_{i=1}^k L_i = \shuffle_{i=1}^k (L_i\bwt i) \ .$$
For example, if we shorten $\tuple{a, 1}$ to $a1$ etc.,
we have
$$\{ab\} \ish \{b\} = \{a1 \, b1\} \shuffle \{b2\} = \{ a1 \, b1 \, b2, a1 \, b2 \, b1,
b2 \, a1 \, b1 \} \ . $$
 It is well known that if $L_i$ is recognized by an NFA
of size $n_i$, then both $\shuffle_{i=1}^k  L_i$ and $\ish_{i=1}^k L_i$
are recognized by NFAs
of size $O(\Pi_{i=1}^k n_i)$.

\section{Models}

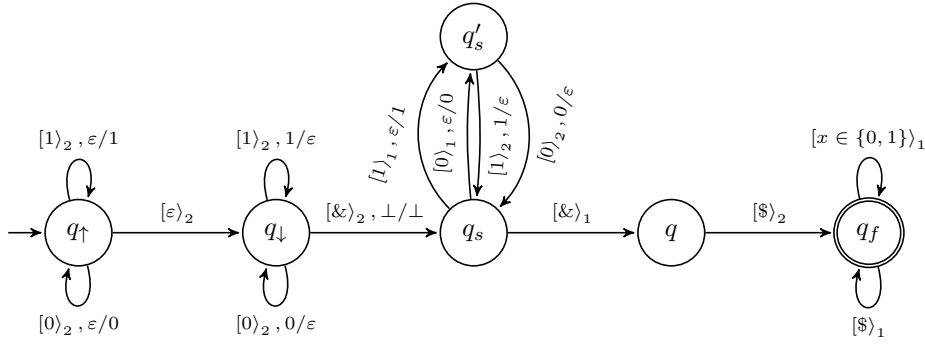
\begin{figure*}
	\centering
	\begin{tikzpicture}[->,>=stealth',shorten >=1pt,auto,node distance=2.6cm, semithick,initial text=]

  \tikzstyle{every state}=[fill=none,draw=black,text=black]

	\node[initial,state]  (Qp)                {$q_{\uparrow}$};
	\node[state]          (Qpr) [right of=Qp] {$q_{\downarrow}$};
  \node[state]          (Qs)  [right of=Qpr]{$q_s$};
  \node[state]          (Qs1) [above of=Qs] {$q_s^\prime$};
  \node[state]          (Qff) [right of=Qs] {$q$};
	\node[accepting,state](Qf)  [right of=Qff]  {$q_f$};

	\path[font=\scriptsize]
	(Qp) edge node[above] {\(\readh{\varepsilon}_2\)} (Qpr)
	(Qp) edge [loop above] node {\(\readh{1}_2,\varepsilon/1\)} (Qp)
	(Qp) edge [loop below] node {\(\readh{0}_2,\varepsilon/0\)} (Qp)
	(Qpr) edge [loop above] node {\(\readh{1}_2,1/\varepsilon\)} (Qpr)
	(Qpr) edge [loop below] node {\(\readh{0}_2,0/\varepsilon\)} (Qpr)
	(Qpr) edge node[above] {\(\readh{\&}_2,\bot/\bot\)} (Qs)
	(Qs) edge [bend left=5,sloped,pos=.9] node {\(\readh{0}_1,\varepsilon/0\)} (Qs1)
	(Qs) edge [bend left=45,sloped,pos=.7] node {\(\readh{1}_1,\varepsilon/1\)} (Qs1)
	(Qs1) edge [bend left=5,sloped,pos=.9] node {\(\readh{1}_2,1/\varepsilon\)} (Qs)
	(Qs1) edge [bend left=45,sloped,pos=.7] node {\(\readh{0}_2,0/\varepsilon\)} (Qs)
	(Qs) edge node[above] {\(\readh{\&}_1\)} (Qff)
	(Qff) edge node[above] {\(\readh{\rmark}_2\)} (Qf)
	(Qf) edge [loop above] node {\(\readh{x \in\set{0,1}}_1\)} (Qf)
	(Qf) edge [loop below] node {\(\readh{\rmark}_1\)} (Qf);
\end{tikzpicture}
\caption{\(2\)-HPDA accepting \(\{ w\& w\mid w\in (\set{0,1})^*\) and \(w\) is a palindrome\(\}\), \(\bot\) is the end-of-stack symbol}
	\label{fig:mhpda_example}
\end{figure*}

A \emph{tape content} (or simply tape) \(w\) over \(\Sigma\) is a word
\(w\in\Sigma^*\).  For \(d\geq 1\), a {\em \(d\)-tuple of tapes} is a
\(d\)-tuple \( (w_1,\dots,w_d) \) where each \(w_i\) is a tape.  Let
\(w\in\Sigma^*\), define \([w]^d\) as the \(d\)-tuple \((w,\dots,w)\). It
extends to languages as follows: let \(L\subseteq \Sigma^*\), we write
\([L]^{d}\) to denote the set of \(d\)-tuples of tapes given by
\(\set{(w_1,\dots,w_d) \mid w_i\in L }\).

\begin{definition}
  A \(d\)-tape pushdown automaton (\(d\)-TPDA, for short) is a 9-tuple
  \(A=\tuple{S,\Sigma,\rmark,\Gamma,M,\nu,s_0,\gamma_0,F}\) where
  \begin{enumerate}
    \item \(S\) is a finite non-empty set of \emph{states},
    \item \(\Sigma\) is the \emph{tape alphabet},
    \item \(\rmark\) is a symbol not in \(\Sigma\) (the \emph{endmarker} for the tape),
    \item \(\Gamma\) is the \emph{stack alphabet},
    \item \(M\), the \emph{set of transitions}, is a mapping from
          \(S\times (\Sigma\cup\set{\rmark}\cup\set{\varepsilon})\times \Gamma\) into the
          finite subsets of \(S\times \Gamma^*\),
    \item \(\nu\colon S\rightarrow[1,d]\) is the \emph{tape selector function},
    \item \(s_0\in S\) is the \emph{start state},
    \item \(\gamma_0\in\Gamma\) is the \emph{initial pushdown symbol},
    \item \(F\subseteq S\) is the set of \emph{final states}.
  \end{enumerate}
\end{definition}

Intuitively, a \(d\)-TPDA has a finite-state control ($S$), $d$ input tapes,
and a stack.
There is a separate input-reading head on each tape.
Each state $s\in S$ in the finite state control reads from the tape given
by $\nu(s)$ and pops the top of the stack.
The transition relation then non-deterministically determines the new control state
and the sequence of symbols pushed on to the stack.
The read head moves one step to the right on its input tape.

For the sake of readability, we write
\((s,\gamma)\trpda{\sigma}(s',w)\) whenever \((s',w)\in M(s,\sigma,\gamma)\).
We sometimes write \((s,\gamma)\trpda{\readh{\sigma}_{i}}(s',w)\) where
\(\nu(s)=i\) when we want to make explicit from which tape we are reading.

The size \(|A|\) of a $d$-TPDA $A$ is given by $|S| + |\Sigma| + |\Gamma| + |M| + |\nu|$,
where in the encoding of the function $\nu$, the numbers in $[1,d]$ are encoded in binary.
Intuitively, \(|A|\) is proportional to the number of bits required to represent
a $d$-TPDA when numbers are represented in binary.

Let us fix a \(d\)-TPDA \(A=\tuple{S,\Sigma,\rmark,\Gamma,M,\nu,s_0,\gamma_0,F}\).

\begin{definition}
  Let \(\#\) be a symbol distinct from symbols in \(\Sigma\cup\set{\rmark}\).
  Define \(\tapes=\set{w \# w' \mid w\cdot
  w'\in \Sigma^* \rmark}\).  An \emph{instantaneous description} (ID) of \(A\)
  is a triple \((s,\tape=\tuple{t_1,\dots,t_d},w)\in S\times [\tapes]^{d} \times \Gamma^*\).
	An ID \((s,\tape,w)\) denotes that \(A\) is in state \(s\),
	with pushdown store content \(w\in \Gamma^*\), and where
	\(\tape=\tuple{t_1,\dots,t_d}\) is such that \(t_i\in\tapes\) gives the
	configuration of tape \(i\) where the position of the head indicated by
	\(\#\).
\end{definition}

\begin{definition}
Let \(\vdash\), be the relation between IDs defined as follows:
let \( c=(s,\tape,w\gamma)\) and \(c'=(s',\tape', w w')\) be two IDs.
We have \(c\vdash c'\) if{}f each of following conditions is satisfied:
  \begin{enumerate}
		\item \((s,\gamma)\trpda{\sigma_r}(s',w')\) for some \(\sigma_r\in\Sigma\cup\set{\varepsilon,\rmark}\).
    \item \(t_{\nu(s)}=x\#\sigma_r y\) and
      \(t'_{i}=
      \begin{cases}
	x\sigma_r\# y &\text{if } i=\nu(s)\\
	t_{i} &\text{else}
      \end{cases}\)
  \end{enumerate}
  \label{def:consecution}
\end{definition}

Let \(\vdash^{*}\) be the reflexive and transitive closure of \(\vdash\).

We now introduce helper functions \(\ltape\) and \(\rtape\) which given
an ID \(c\) and a tape \(h\in [1,d]\) returns the tape content
lying to the left and to the right (without the head position), respectively.
\begin{definition}
	Given an ID \(c=(s,\tape=\tuple{t_1,\dots,t_d},w)\) and \(h\in[1,d]\), define \(\rtape(c,h)\) and
	\(\ltape(c,h)\) as follows: let \(t_h=w_1\# w_2\) then \(\rtape(c,h)= w_2\)
	and \(\ltape(c,h)=w_1\).
\end{definition}

Let us now define the languages accepted by \(d\)-TPDA.
\begin{definition}
	Given a ID \(c\), we say that a head \(i\) is \emph{off its tape } in \(c\) whenever \(\rtape(c,i)=\varepsilon\).
	Let \(c=(s,\tape,w)\) be an ID, we say that \(c\) is
	\emph{accepting} if{}f \(s\in F\) and for every \(i\in[1,d]\) head \(i\)
	is off its tape in \(c\).
	A \(d\)-tuple of tapes \( (x_1,\dots,x_d)\in[\Sigma^*]^{d}\) is
	\emph{accepted} by \(A\) if
	\((s_0,\tuple{\#x_1\rmark,\dots,\# x_d\rmark},\gamma_0)\vdash^{*}
	(s,\tape,w)\) for some ID \((s,\tape,w)\) that is accepting. The set of
	\(d\)-tuple of tapes accepted by \(A\) is denoted \(T(A)\).  A subset \(L\subseteq
	[\Sigma^*]^{d}\) is \(d\)-TPDA \emph{definable} if there exists some
	\(d\)-TPDA \(A\) such that \(L=T(A)\).
\end{definition}

\begin{remark}
\label{rmk:generalremarks}%
\hspace{0pt}
\begin{itemize}
\item  Having all heads off their tape is a necessary condition to accept.
		Therefore any accepting run (even if the tape is \([\varepsilon]^d\)) needs to
		perform at least one read on each tape because of \(\rmark\).
    This implies that for any non-trivial $d$-TPDA, $d \leq |S|$.
\item The language of \(d\)-TPDA are recursive for each $d > 0$ \cite{IbarraPhD67}.
    The languages that are \(1\)-TPDA definable are the CFLs.
    Observe the following difference w.r.t. classical definition, e.g. as in \cite{HU79}.
    In fact, for a \(1\)-TPDA to accept we need the current control state to be final and the head to be off the tape.
\end{itemize}%
\end{remark}

In latter sections, we use a graphical notation for MHPDAs because it better
carries intuitions.  Fig.~\ref{fig:mhpda_example} gives such an example of
\(2\)-HPDA which recognize a language over symbols \(\set{0,1,\&}\) given by
\(\{ w\& w\mid w\in (\set{0,1})^*\) and \(w\) is a palindrome\(\}\).
Intuitively, in \(q_{\uparrow}\) the \(2\)-HPDA uses head \(2\) to recognize
the first palindrome using its stack. When head \(2\) reads \(\&\) the
MHPDA enters \(q_s\) where it checks using both heads that the subwords before
and after \(\&\) are identical. If the check succeeds then the MHPDA enters
\(q\) then \(q_f\) (head \(2\) has fallen off the tape) where it accepts after
making head \(1\) fall off the tape.  The transition from $q_{\downarrow}$ to
\(q_s\) labelled $\readh{\&}_2, \bot / \bot$ reads as follows: if in state
$q_{\downarrow}$ stack symbol \(\bot\) is on the top of the stack then read $\&$ with head
\(2\) and update the location to $q_s$.  Also read the transition from $q_f$ to
itself and labelled $\readh{x\in\set{0,1}}_1$ as follows: in state $q_f$ read
any symbol of \(\set{0,1}\), go to $q_f$.  In what follows, to ease the
readability we omit the formal description of the automata and use our
graphical notation instead.

We now introduce a generalization of pushdown automata with several
{\em heads} working on a shared tape.
This model is closely related to $d$-TPDA as described below.

\begin{definition}
	Let \(\Delta_d\subseteq [\Sigma^*]^{d}\) be given by
	\(\set{(w_1,\dots,w_d)\in[\Sigma^*]^{d}\mid w_1=\dots=w_d}\) and \(\proj_1\colon [\Sigma^*]^d \rightarrow \Sigma^*\) to
	be the function which maps \(L\subseteq [\Sigma^*]^{d}\) onto the
	first tape: \(\proj_1(L)=\set{ w_1\in\Sigma^* \mid (w_1,\dots,w_d)\in L}\).
	\label{def:delta}
\end{definition}

When the \(d\)-tuple of tapes is restricted to \(\Delta_d\), that is, when all the
tapes have identical content, we can view \(A\) as a pushdown automaton with
\(d\)-heads sharing a unique tape.
In this case we define the language
\(L\subseteq \Sigma^*\) accepted by the \(d\)-head pushdown automaton \(A\)
(or \(d\)-HPDA) to be \(\proj_1(T(A)\cap \Delta_d)\)
and we denote this language by \(L(A)\).
We write MHPDA for the class of models $d$-HPDA for $d\geq 1$.

\section{Emptiness modulo Bounded Expressions}
\label{sec:decisionprocedure}

Given a $d$-HPDA $M$ and
a bounded expression $\pat=w_1^* \ldots w_n^*$, both over an alphabet $\Sigma$,
we show how to check emptiness of $L(M) \cap \pat$. Recall that we can
construct a $d$-TPDA $A$ of size $O(|M|)$ such that $L(M) \cap \pat = \varnothing$ if{}f $T(A) \cap [\pat]^d \cap \Delta_d=\varnothing$, where
$\Delta_d$ is the set of $d$-tuples of the form $(w, w, \ldots, w)$ (see def.~\ref{def:delta}).

\begin{figure}[!b]
\centering
\begin{tikzpicture} \matrix(m)[matrix of math nodes, row sep=.7em, column sep=3.3em, text height=1.5ex, text depth=0.25ex]
       {M&A&B&C&G&\Phi\\ \pat & & & & & \\};
       \path[->,font=\scriptsize,>=angle 90]
			 (m-2-1.south) edge[densely dotted,bend right=5,in=210] node[above]{\(O(|M^3|\cdot |\pat|^{3d})\)}
			 				 																node  [below]{Thm.~\ref{thm:emptiness}} 		(m-1-6.south)
       (m-1-1) edge node[above]{\(O(|M|)\)} (m-1-2)
       (m-1-2) edge node[above]{\(O(|A|\cdot |\pat|^d)\)}
                    node(label)[below]{Lem.~\ref{lem:lemmaone}} (m-1-3)
			 (m-1-3) edge node[above]{\(O(|B|)\)}
									  node[below]{Lem.~\ref{lem:lemmatwo}} (m-1-4)
			 (m-1-4) edge node[above]{\(O(|C|^3)\)}
									  node[below]{Lem.~\ref{lem:grammars}} (m-1-5)
			 (m-1-5) edge node[above]{\(O(|G|)\)}
			 							node[below]{Thm.~\ref{thm:parikh-presburger}} (m-1-6);
		\draw[->,>=angle 90] (m-2-1) -| (label.south);
		\draw[thick,dotted] (m-1-1.north west)  rectangle (m-2-1.south east);
\end{tikzpicture}
\caption{Summary of the decision procedure steps}
\label{fig:decproc}
\end{figure}
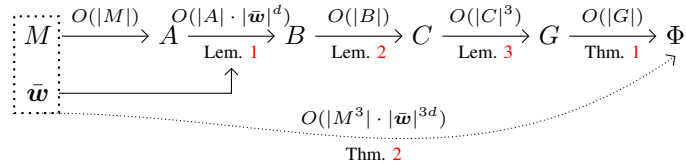

In Section \ref{subsec:emptiness1} we show that emptiness of $T(A) \cap [\pat]^d$
can be reduced to emptiness of a context-free grammar, and in Section \ref{subsec:emptiness2} that emptiness of $L(M) \cap \pat$ can be reduced to unsatisfiability of an existential Presburger formula.
The steps of the reduction are summarized in Fig.~\ref{fig:decproc}.

\subsection{Emptiness of $T(A) \cap [\pat]^{d}$}
\label{subsec:emptiness1}
We construct in three steps a context-free grammar that recognizes an ``encoding'' of  $T(A) \cap [\pat]^d$.

Roughly speaking, in the first step we construct a $d$-TPDA recognizing the
result of applying a transformation on $T(A) \cap [\pat]^d$ which ``contracts''
each word $w_i$ of $\pat$ into a single letter.

Let $\oSigma =\{a_1, \ldots, a_n\}$ be a new alphabet and let \(\pata=a_1^*\cdots a_n^*\) be
a bounded expression over \(\oSigma\). Given a bounded expression \(\pat=w_1^*\cdots w_n^*\) over \(\Sigma\), we
define the mapping \(f_{\pat}\colon \nats^{n}\rightarrow \Sigma^*\) by
\(f_{\pat}\colon (i_1,\ldots,i_n)\mapsto w_1^{i_1}\cdots w_n^{i_n} \).

\begin{lemma}
 There is a computable $d$-TPDA $B$ over $\oSigma$ of
 size $O(|A| \cdot |\pat|^d)$ such that for every \(\boldsymbol{k_1},\ldots,\boldsymbol{k_d}\in\nats^n\)
 we have:
 \begin{gather*}
	 \bigl(f_{\pat}(\boldsymbol{k_1}),\ldots,f_{\pat}(\boldsymbol{k_d})\bigr) \in T(A)\cap [\pat]^{d}\\
	 \text{if{}f}\\
	 \bigl(f_{\pata}(\boldsymbol{k_1}),\ldots,f_{\pata}(\boldsymbol{k_d})\bigr) \in T(B)\enspace .
 \end{gather*}
 \label{lem:lemmaone}
\end{lemma}
\begin{IEEEproof}[Sketch of Proof] We first construct a $d$-TPDA $B_1$ such that \(T(B_1) = T(A) \cap
[\pat]^d\).  For this, let $W$ be an NFA recognizing $\pat\cdot\rmark$, and let
$Q_W$ be its set of states and \(F_W\subseteq Q_W\) the accepting ones. Adapting the shuffle construction for
NFAs, we can construct a NFA $W^d$ with states $[Q_W]^d = \underbrace{Q_W
\times \cdots \times Q_W}_{\mbox{\scriptsize $d$-times}}$ recognizing
\(\ish_{i=1}^d L(\pat\cdot \$)\).  We synchronize $A$ with $W^d$ as follows.
The set of states of $B_1$ is $S \times [Q_W]^d$, where $S$ is the set of
states of $A$, and the set of final states is $F \times [F_W]^d$.
The tape selection function of \(B_1\) is determined by the one of \(A\).
If $A$ has a
transition $(s_a,\gamma) \trpda{\readh{\sigma}_{\ell}}(s_b,w)$, where
$\sigma\in\Sigma\cup\set{\rmark}$ is read from the tape $\ell=\nu(s_a)$, and \(W^d\) has a
transition \(\tuple{q_1, \ldots, q_\ell, \ldots,
q_d}\stackrel{\tuple{\sigma,\ell}}{\rightarrow} \tuple{q_1, \ldots, q'_\ell,
\ldots, q_d}\), then $B_1$ has a transition \((\tuple{s_a,q_1, \ldots, q_\ell,
\ldots, q_d}, \gamma)\trpda{\readh{\sigma}_{\ell}}(\tuple{s_b,q_1, \ldots,
q_\ell', \ldots, q_d},w)\).
If \(A\) has a
transition $(s_a,\gamma) \trpda{\readh{\varepsilon}_{\ell}}(s_b,w)$
(resp. \(W^d\) has a
transition \(\tuple{q_1, \ldots, q_j, \ldots,
q_d}\stackrel{\tuple{\varepsilon,j}}{\rightarrow} \tuple{q_1, \ldots, q'_j,
\ldots, q_d}\)), then $B_1$ has transition \((\tuple{s_a,q_1, \ldots, q_d}, \gamma)\trpda{\readh{\varepsilon}_{\ell}}(\tuple{s_b,q_1, \ldots,q_d},w)\)
(resp. \((\tuple{s_a,q_1, \ldots, q_j,
\ldots, q_d}, \gamma)\trpda{\readh{\varepsilon}_{\ell}}(\tuple{s_a,q_1, \ldots,
q'_j, \ldots, q_d},\gamma)\) for every \(\gamma\in \Gamma\)).
$B_1$ has no further transitions.

Now we construct $B$. It is easy to construct $W$ so that for every  word $w_i$
of $\pat$ it contains a state $q_{w_i}$ that is entered every time (and only
when) $W$ reads the last letter of $w_i$. We proceed as follows. First, we
transform all transitions of $B_1$, with the exception of those labeled with
endmarkers, into $\varepsilon$-transitions. Then, we relabel again all
transitions entering $q_{w_i}$, i.e, all transitions in which some copy of $W$
takes a transition with target $q_{w_i}$: we replace $\varepsilon$ by $a_i$.
\end{IEEEproof}

In a second step we construct a PDA that recognizes the indexed shuffle of $T(B)$.
Let $\oSigma_d = \bigcup_{i=1}^d (\oSigma \bwt i)$.
Given a \(d\)-tuple of tapes \(u=(u_1,\ldots,u_d)\) define \(\ish(u)=\ish_{i=1}^{d} \set{u_i}\).

\begin{lemma}
There is a computable PDA $C$ over $\oSigma_d$ of size $O(|B|)$
such that $u \in T(B)$ if{}f $\ish(u) \cap L(C) \neq \varnothing$ for every $u \in [\oSigma^*]^d$.
\label{lem:lemmatwo}
\end{lemma}
\begin{IEEEproof}[Sketch of Proof] $B$ and $C$ have the same states, initial
and final states, and stack alphabets.  Assume $B$ is currently at state $s$,
and the tape selector assigns to $s$ tape number $\ell=\nu(s)$.  The
transitions of $C$ are defined so that if in the next move $B$ reads a letter
$\sigma$, then $C$ reads the letter $\tuple{\sigma,\ell}$ (unless
$\sigma\in\set{\rmark,\varepsilon}$, in which case $C$ reads $\varepsilon$).
Formally, for $ \sigma \neq \rmark$ and \(\sigma\neq\varepsilon\) the PDA $C$
has a transition \((s,\gamma)\trpda{\tuple{\sigma,\ell}}(s',w)\) if{}f $B$ has
a transition \( (s,\gamma) \trpda{\readh{\sigma}_{\ell}}(s',w)\), and $C$ has a
transition \((s,\gamma)\trpda{\varepsilon}(s',w)\) if{}f $B$ has a transition
\( (s,\gamma) \trpda{\readh{\rmark}_{\ell}}(s',w)\) or \( (s,\gamma)
\trpda{\readh{\varepsilon}_{\ell}}(s',w)\).
Now, $C$ accepts the word of $\ish(u)$ that interleaves the letters from the
different tapes in the order in which they are read by $B$.
\end{IEEEproof}

The third step is standard \cite{HU79}:

\begin{lemma}
There is a computable CFG \(G\) over $\oSigma_d$
of size $O(|C|)^3$ such that  \(L(G)=L(C)\).
\label{lem:grammars}
\end{lemma}

Putting these lemmas together, we finally get

\begin{proposition}
\label{prop:summ}
There is a computable CFG \(G\) over $\oSigma_d$ of size $O(|A|^3 \cdot |\pat|^{3d})$
such that for every \(\boldsymbol{k_1},\ldots,\boldsymbol{k_d}\in\nats^n\) we have:
 \begin{gather*}
	 \bigl(f_{\pat}(\boldsymbol{k_1}),\ldots,f_{\pat}(\boldsymbol{k_d})\bigr) \in T(A)\cap [\pat]^{d}\\
	 \text{if{}f}\\
	 \ish\bigl(f_{\pata}(\boldsymbol{k_1}),\ldots,f_{\pata}(\boldsymbol{k_d})\bigr)\cap L(G)\neq\varnothing
 \end{gather*}
\end{proposition}

\subsection{Emptiness of $L(M) \cap \pat$}
\label{subsec:emptiness2}

Recall that $L(M) \cap \pat = \varnothing$ if{}f $T(A) \cap [\pat]^{d} \cap
\Delta_d = \varnothing$. To decide this problem, we rely on the notion of
Parikh image.  By definition of indexed shuffle, for every tuple $v \in
[\oSigma^*]^d$ all the words of $\ish(v)$ have the same Parikh image, which
justifies the notation $\parikh(\ish(v))$. Now we have:

\begin{lemma}
For every $v \in [\oSigma^*]^d$:
$\ish(v) \cap L(G) \neq \varnothing$ if{}f $\parikh(\ish(v)) \in \parikh(L(G))$.
\label{lem:parikhreasoning}
\end{lemma}
\begin{IEEEproof}
The right-to-left direction is obvious. For the converse, $\ish(v) \cap L(G) \neq \varnothing$ implies $\parikh(v') \in \parikh(L(G))$ for some $v' \in \ish(v)$,
but all elements of $\ish(v)$ have the same Parikh mapping.
\end{IEEEproof}

So checking $\ish(v) \cap L(G) \neq \varnothing$ can be done by checking
$\parikh(\ish(v)) \in \parikh(L(G))$. For this check we can resort to the following theorem.

\begin{theorem}{\cite{Seidl05}}
\label{thm:parikh-presburger}
For each CFG $G$, there is a computable existential Presburger formula \(\Phi\) of size $O(|G|)$
such that  \(\parikh(L(G)) = \denote{\Phi}\).
\end{theorem}

We immediately get:

\begin{proposition}
\label{prop:formulaphi}
There is a computable existential Presburger formula $\Phi$ with free
variables $\{x_{ij} \mid i\in [1,n], j\in [1,d]\}$ of size $O(|G|)$ such that
\begin{gather*}
	\bigl(f_{\pat}(\boldsymbol{k_1}),\ldots,f_{\pat}(\boldsymbol{k_d})\bigr) \in T(A) \cap [\pat]^d\\
	\text{if{}f} \\
	\Phi(\boldsymbol{k_1},\ldots,\boldsymbol{k_d})\text{ is true}\enspace .
\end{gather*}
\end{proposition}%
\begin{IEEEproof}  Take for $\Phi$ the formula of Thm.~\ref{thm:parikh-presburger}. We have:
\begin{align*}
	& \bigl(f_{\pat}(\boldsymbol{k_1}),\ldots,f_{\pat}(\boldsymbol{k_d})\bigr) \in T(A) \cap [\pat]^d\\
	\text{if{}f }& \ish\bigl(f_{\pata}(\boldsymbol{k_1}),\ldots,f_{\pata}(\boldsymbol{k_d})\bigr)\cap L(G)\neq\varnothing & \mbox{Prop.~\ref{prop:summ}} \\
	\text{if{}f }& (\boldsymbol{k_1},\ldots,\boldsymbol{k_d}) \in \parikh(L(G)) &\mbox{Lem.~\ref{lem:parikhreasoning}} \\
	\text{if{}f }& \Phi(\boldsymbol{k_1},\ldots,\boldsymbol{k_d}) \text{ is true} & \mbox{Thm.~\ref{thm:parikh-presburger}}
\end{align*}%
\end{IEEEproof}

The advantage of Prop.~\ref{prop:formulaphi} is that it can be easily extended to a procedure for checking not only emptiness of $T(A) \cap [\pat]^d$, but also emptiness of $T(A) \cap [\pat]^d \cap \Delta_d$. Recall that the tuples in $T(A) \cap [\pat]^d \cap \Delta_d$ are the tuples of $T(A)$ of the form
$(w, \ldots, w) \in [\Sigma^*]^d$ for some $w \in \pat$.  Let \(\boldsymbol{k},\boldsymbol{k_1},\dots,\boldsymbol{k_d}\in\nats^n\).  We have:
\begin{align*}
	&\bigl(f_{\pat}(\boldsymbol{k_1}),\ldots,f_{\pat}(\boldsymbol{k_d})\bigr) \in T(A)\cap[\pat]^d\cap\Delta_d\\
	\text{if{}f } &( \text{property of } \Delta_d, , \boldsymbol{k}=\boldsymbol{k_1}=\cdots=\boldsymbol{k_d})\\
	&[f_{\pat}(\boldsymbol{k})]^d \in T(A)\cap[\pat]^d\\
	\text{if{}f }& (\text{Prop.~\ref{prop:formulaphi}})\\
	&\Phi(\underbrace{\boldsymbol{k},\ldots,\boldsymbol{k}}_{d \text{ times}}) \text{ is true}\\
	\text{if{}f }& \exists \boldsymbol{i_1},\ldots,\boldsymbol{i_d}\in\nats^{n}\colon\Phi(\boldsymbol{i_1},\ldots,\boldsymbol{i_d}) \text{ is true}\\
	&\text{and }  \boldsymbol{i_1}=\cdots=\boldsymbol{i_d}\\
	\text{if{}f }& \exists x_{11} \ldots \exists x_{kd}  \left(\Phi \wedge \textstyle{\bigwedge_{i=1}^n} \textstyle{\bigwedge_{j=1}^{d}} x_{ij}=k_{i} \right)\text{ is true}
\end{align*}
\vspace{0.2cm}
\noindent  where $\Phi$ is the formula of Prop.~\ref{prop:formulaphi}. So we get

\begin{theorem}{}
\label{thm:emptiness}
There is a computable formula $\Psi(x_1, \ldots, x_n)$ of existential Presburger arithmetic of size $O(|M|^3 \cdot |\pat|^{3d})$ such that $f_{\pat}(k_1,\ldots,k_n) \in L(M) \cap \pat$
if{}f  $\Psi(k_{1}, \ldots, k_{n})$ is true. In particular, $L(M) \cap \pat \neq \varnothing$ if{}f $\Psi$ is satisfiable.
\end{theorem}
\begin{IEEEproof}
It suffices to take $\Psi (x_1, \ldots, x_n) = \exists x_{11} \ldots \exists x_{kd}  \colon \left( \Phi \wedge \bigwedge_{i=1}^n \bigwedge_{j=1}^{d} x_{ij}=x_{i} \right)$.
\end{IEEEproof}

This  theorem admits a simple but useful generalization:
\begin{theorem}{}
	Let \(\set{M_i}_{i\in [1,q]}\) be a family of MHPDA such that \(M_i\) is a \(c_i\)-HPDA for each \(i\in[1,q]\).
	Let \(c=\max\bigl(\set{c_i}_{i\in [1,q]}\bigr)\) and \(m=\max\bigl(\set{|M_i|}_{i\in [1,q]}\bigr)\).
	There is a computable formula $\Psi(x_1, \ldots, x_n)$ of existential Presburger arithmetic of size $O( q\cdot m^3 \cdot |\pat|^{3c})$ such that $f_{\pat}(k_1,\ldots,k_n) \in \bigcap_{i=1}^q L(M_i) \cap \pat$
	if{}f  $\Psi(k_{1}, \ldots, k_{n})$ is true.
	\label{thm:superemptiness}
\end{theorem}
\begin{IEEEproof}
	Define \(\Psi(x_1,\dots,x_n)\) to be \(\textstyle{\bigwedge_{i=1}^q} \Psi_{i}(x_1,\dots,x_n)\) such that
	each \(\Psi_{i}(x_1,\dots,x_n)\) is the formula obtained by Thm.~\ref{thm:emptiness} on input \(M_i\) and \(\pat\).
	Correctness is proved as follows:
	\begin{align*}
		&f_{\pat}(k_1,\ldots,k_n) \in \textstyle{\bigcap_{i=1}^q} L(M_i) \cap \pat\\
		\text{if{}f } &\textstyle{\bigwedge_{i=1}^q} f_{\pat}(k_1,\ldots,k_n)\in L(M_i)\cap\pat\\
		\text{if{}f } &\textstyle{\bigwedge_{i=1}^q} \Psi_{i}(x_1,\dots,x_n) \text{ is true} &\text{Thm.~\ref{thm:emptiness}}\\
		\text{if{}f } & \Psi(x_1,\dots,x_n) \text{ is true}&\text{def.\ of }\Psi
	\end{align*}
	We conclude from Thm.~\ref{thm:emptiness} that \(|\Psi_i|=O(m\cdot |\pat|^{3c})\) for each \(i\in[1,q]\), hence
	that \(|\Psi|=O(q\cdot m\cdot |\pat|^{3c})\).
\end{IEEEproof}

\subsection{Complexity}

Emptiness of MHPDAs is clearly undecidable (by reduction from the emptiness
problem for intersection of context-free languages).
We prove that emptiness modulo a bounded expression is coNEXPTIME-complete.

\begin{theorem}{}
The emptiness problem for MHPDAs modulo an arbitrary bounded expression is
in coNEXPTIME. Moreover, the emptiness problem for MHPDAs
and $\pat=(01)^*$ is coNEXPTIME-hard.
\label{prop:conexphard}
\end{theorem}

The question arises whether emptiness remains coNEXPTIME-complete for letter-bounded expressions.
Remarkably, this is not the case:
for such expressions the emptiness problem is only NP-complete.
Fix a letter-bounded expression $\patb = b_1^* \ldots b_n^*$ where \(b_i\)'s are not
necessarily distinct. The key to the result is that Lem.~\ref{lem:lemmaone}
(with \(\pat\) now equal to \(\patb\)) can be replaced by the following one.

\begin{lemma}
There is a family $\{B_i\}_{i=1}^\alpha$ of $d$-TPDAs over $\oSigma$,
where $\alpha = d^{|\patb| d}$ and each $B_i$ has size $O(|A| \cdot |\patb|\cdot d^{2})$,
 such that for every
\(\boldsymbol{k_1},\ldots,\boldsymbol{k_d}\in\nats^n\)
 we have
 \begin{gather*}
	 \bigl(f_{\patb}(\boldsymbol{k_1}),\ldots,f_{\patb}(\boldsymbol{k_d})\bigr) \in T(A)\cap [\pat]^{d}\\
	 \text{if{}f}\\
	 \bigl(f_{\pata}(\boldsymbol{k_1}),\ldots,f_{\pata}(\boldsymbol{k_d})\bigr) \in \textstyle{\bigcup_{i=1}^{\alpha}}  T(B_i)
 \end{gather*}
\noindent Moreover, we can decide in time $O(|A| \cdot |\patb|\cdot d^2)$ if a given MHPDA belongs to
$\{B_i\}_{i=1}^\alpha$.
 \label{lem:lemmaNP}
\end{lemma}
\begin{IEEEproof}
We can easily construct an NFA $W$ recognizing $L(\patb\cdot \rmark)$ with
states $\{q_1, \ldots, q_{n+1}\}$ (recall that $n+1 = |\patb|$), initial state $q_1$,
final state $q_{n+1}$, and transitions \(\set{q_i \stackrel{b_i}{\rightarrow} q_i
\mid i\in[1,n]}\) \(\cup\) \(\set{q_{j}\stackrel{\varepsilon}{\rightarrow}
q_{j+1}\mid j\in[1,n-1]}\) \(\cup\) \(\set{q_n\stackrel{\rmark}{\rightarrow}
q_{n+1}}\). Let $W^d$ be the NFA defined in Lem.~\ref{lem:lemmaone} recognizing
\(\ish_{i=1}^d L(\patb\cdot \rmark)\).
While $W^d$ has $(n+1)^d$ states,
it is easy to see that for $\pat=\patb$ every accepting run of $W^d$ only
visits $(n+1)\cdot d$ {\em distinct} states, because every transition
 $\tuple{q_{i_1}, \ldots, q_{i_d}} \stackrel{\tuple{\sigma,\ell}}{\rightarrow} \tuple{q_{j_1}, \ldots, q_{j_d}}$  of $W^d$ satisfies $i_1 \leq j_1, \ldots, i_d
\leq j_d$
We can then associate to each accepting run $\rho$ the subset
\(Q^{\rho}_{W^d}\) of the states of \(Q_{W^d}\) visited by \(\rho\), and so the
sub-NFA $W^d_\rho$ of $W^d$ with \(Q^{\rho}_{W^d}\) as set of states, and whose
transitions are the transitions of $W^d$ between states of \(Q^{\rho}_{W^d}\).
Clearly, $W^d_\rho$ has at most \( (n+1)\cdot d\) states and at most \(( (n+1)\cdot
d)\cdot(d+d)=O(n\cdot d^2)\) transitions. (Let a state \(\tuple{q_{i_1},\dots, q_{i_d}}\); the term \( (d+d)\) corresponds to the
transitions labeled by \(\tuple{b_{i_j},j}\) or \(\tuple{\varepsilon,j}\) for each \(j\in [1,d]\).)
Moreover, even though there are
infinitely many accepting runs, the number of different such sub-NFAs is
$d^{|\patb|d}$, because each state of $W^d$ has $d$ successors different from itself,
and every accepting run of $W^d$ only visits $ (n+1)\cdot d$ distinct states.
Let $W^d_1, \ldots, W^d_\alpha$ be an enumeration of them.

In Lem.~\ref{lem:lemmaone} we first construct a $d$-TPDA $B_1$ by
synchronizing $A$ and $W^d$, and then we transform $B_1$ into another $d$-TPDA
$B$.  Now we first synchronize $A$ and $W^d_i$, yielding a $d$-TPDA $B_{1i}$
for every $i\in [1, \alpha]$, and then we apply the same transformation as in
Lem.~\ref{lem:lemmaone} to obtain a $d$-TPDA $B_i$. Clearly, we have $T(B) =
\bigcup_{i=1}^\alpha T(B_i)$, and so the result follows.
\end{IEEEproof}

Proceeding as in the previous section, we now obtain for each $d$-TPDA $B_i$
a grammar $G_i$, and from it an existential Presburger formula $\Psi_i$.
We get:

\begin{proposition}{}
\label{prop:emptinessNP}
There is a computable family $\{\Psi_i(x_1, \ldots, x_n)\}_{i=1}^\alpha$ of existential Presburger formulas,
each of them of size $O(|M|^3 \cdot|\patb|^3 \cdot d^6)$, such that $f_{\patb}(k_1,\ldots,k_n) \in L(M) \cap \patb$
if{}f  $\bigvee_{i=1}^\alpha \Psi_i(k_{1}, \ldots, k_{n})$ is true. In particular, $L(M) \cap \patb \neq \varnothing$ if{}f
at least one of the formulas in the family is satisfiable.
Moreover, we can decide in time $O(|M|^3 \cdot|\patb|^3 \cdot d^6)$ if a given formula belongs to
$\{\Psi_i(x_1, \ldots, x_n)\}_{i=1}^\alpha$
\end{proposition}
Finally, we get:

\begin{theorem}{}
The emptiness problem for MHPDAs modulo letter-bounded expressions
is in coNP.
Moreover, the emptiness problem for MHPDAs and $\pat=b^*$ is coNP-hard.
\end{theorem}
\begin{IEEEproof}
Let $M$ be a $d$-HPDA and let $\patb$ be a letter-bounded expression.
The nondeterministic polynomial algorithm for non-emptiness of $L(M) \cap \patb$ first
guesses one of the formulas $\Psi_i$ of Prop.~\ref{prop:emptinessNP}, checks in polynomial
time that it belongs to the family and then nondeterministically checks that it is
satisfiable. Since $\Psi_i$ has polynomial size in $|M|+|\patb|+d$, the whole
procedure takes nondeterministic polynomial time.

The coNP-hardness result follows from \cite[Theorem 1]{ge11}, which proves that
given CFGs $G_1, \ldots, G_k$, deciding non emptiness of $L(G_1) \cap \ldots \cap L(G_k) \cap L(b^*)$
is coNP-hard. Since we can easily construct in linear time a $k$-HPDA recognizing
$L(G_1) \cap \ldots \cap L(G_k)$, the result follows.
\end{IEEEproof}

\section{Closure under Boolean operations}

It is straightforward to show that MHPDAs are effectively closed under union
and intersection.

\begin{proposition}
Let $A_1$ be a $k_1$-HPDA and $A_2$ a $k_2$-HPDA. We can construct in linear time
 \((k_1+k_2)\)-HPDAs $A_\cup$ and $A_\cap$ such that $L(A_\cup)=L(A_1)\cup L(A_2)$ and
$L(A_\cap)=L(A_1)\cap L(A_2)$.
\label{lem:kcmcapb}
\end{proposition}
\begin{IEEEproof}
$A_\cup$ nondeterministically decides to simulate $A_1$
or $A_2$; it requires $\max{k_1,k_2}$ heads.
$A_\cap$ simulates $A_1$ with heads $[1,k_1]$ and if $A_1$
reaches an accepting state, then it simulates $A_2$ with
heads $[k_1+1,k_1+k_2]$.
\end{IEEEproof}

MHPDAs are not closed under complement, but closed under complement modulo
any bounded expression.

\begin{proposition}
Given an $d$-HPDA \(A\) and a bounded expression \(\pat\),
there is an MHPDA \(B\) such that \(L(B)=\pat\setminus L(A)\) and
\(|B|\) is at most triply exponential in \(|A|\), \(|\pat|\).
\label{prop:comp}
\end{proposition}
\begin{IEEEproof}
The complementation procedure works as follows:
\begin{itemize}
\item Compute the existential Presburger formula $\Psi$ of Thm.~\ref{thm:emptiness} with constants written in unary. A simple inspection of the result of \cite{Seidl05} shows that the size of $\Psi$ is still $O(|A|^3 \cdot |\pat|^{3d})$
(the constants of $\Phi$ for a context-free grammar $G$ have linear size in
$|G|$ even when written in unary).
\item Compute a quantifier-free formula $\Phi \equiv \neg \Psi$ (with constants written in unary). This is possible
because Presburger arithmetic has quantifier elimination procedures. Moreover,
since $\Psi$ has one single block of existential quantifiers, we have
$|\Phi| \in {\it 2exp}(O(|\Psi|))$
\cite{DBLP:conf/stoc/ReddyL78}\cite{DBLP:journals/tcs/Furer82},
where ${\it 2exp}(n) = 2^{2^n}$.
We have $w_1^{k_1}\dots w_n^{k_n}\in (\pat\setminus L(A))$ if{}f $\Psi(k_1, \ldots, k_n)$ is false
if{}f  $\Phi(k_1, \ldots, k_n)$ is true.
\item
Construct the MHPDA $B$ as follows. $B$ has a head for each atomic formula
of $\Phi$. Control ensures that heads read the input one after the other
(i.e., the $i+1$-st head starts reading the input after the $i$-th head has completely read it). The $i$-th head checks whether the $i$-th atomic formula is
satisfied by the input. For instance, a constraint like $3k_1 -2k_2\leq 5$ is checked using the stack as follows:
the stack is used as a counter over the integers
(using two symbols, say $P$ and $N$, and encoding $i$ where \(i\geq 0\) as $P^i
\bot$ and $-i$ (\(i>0)\) as $N^i \bot$ for some bottom stack symbol $\bot$);
$B$ reads $w_1^{k_1}w_2^{k_2}$, so that at the end the counter contains $3k_1
-2k_2$; then $B$ compares the content of the counter with $5$. Control takes
care of evaluating the formula by combining the results of the evaluation of
the atomic formulas. $B$ accepts $w_1^{k_1}\ldots w_n^{k_n}$ if the evaluation
of $\Phi$ is true. Since the constants of $\Phi$ are written in unary, we have
$|B| \in O(|\Phi|)$.  \end{itemize} This procedure yields a triple exponential
bound for $B$ in the size of $A$.  More precisely, the procedure is only triply
exponential in the number of heads of $A$, but not on its number of states or
transitions.
\end{IEEEproof}

For letter-bounded expressions, we get one exponential less by using Prop.~\ref{prop:emptinessNP}
to compute a family of exponentially many Presburger formulas, each
polynomial in the size of the automaton and the bounded expression, then following
the previous construction and noting that the intersection
of exponentially many MHPDAs, each doubly exponential, still gives
a doubly exponential MHPDA.

\begin{proposition}
Given a $d$-HPDA \(A\) and a letter-bounded expression \(\patb\),
there is an MHPDA \(B\) such that \(L(B)=\patb\setminus L(A)\) and
\(|B|\) is at most doubly exponential in \(|A|\), $|\patb|$.
\label{prop:letter-bounded-complement}
\end{proposition}

\section{Optimality questions}
Let ${\cal P}$ denote the class of finite unions of bounded expressions, let ${\cal F}$ denote the
class of finite languages, and let ${\cal U} = {\cal P} \cup {\cal F}$. We have shown that MHPDA is perfect
modulo ${\cal U}$. This raises two questions: (1) is MHPDA perfect modulo some class of regular languages
larger than ${\cal U}$?, and (2) is some class larger than MHPDA perfect modulo ${\cal U}$?.

Prop.~\ref{prop:maxCSL}.1 shows that the answer to (1) is negative. We do not settle (2), but
show in Prop.~\ref{prop:maxCSL}.2 that the largest class of regular languages
for which  the context-sensitive languages (CSL) are perfect is ${\cal F}$. Actually, the
proposition shows that no class with an undecidable emptiness problem (and satisfying some additional very weak
properties) can be perfect modulo any class of regular languages larger than ${\cal F}$. So, in particular,
no class containing the languages generated by Okhotin's conjunctive grammars can be perfect \cite{Okhotin00}.

\begin{proposition}
\label{prop:maxCSL}
\hspace{0pt}
\begin{enumerate}
	\item ${\cal U}$ is the largest class of regular languages such that MHPDA is perfect modulo ${\cal U}$;
	\item ${\cal F}$ is the largest class of regular languages such that CSL is perfect modulo ${\cal F}$.
\end{enumerate}
\end{proposition}

\begin{IEEEproof}
\hspace{0pt}

\noindent
{\bf point 1.}
Let ${\cal C }$ be a class of regular languages stronger than ${\cal U}$.
We show that the emptiness problem of MHPDA modulo ${\cal C}$ is undecidable,
which implies that MHPDA is not perfect modulo ${\cal C}$.

Since ${\cal C }$ is stronger than ${\cal U}$, there is an infinite regular language
$L \in {\cal C}$ that is not equal to a finite union of bounded expressions.
We show that there are words $u, v_0, v_1, x$, such that
$\varepsilon \neq v_0 \neq v_1 \neq \varepsilon$, \(v_0v_1\neq v_1v_0\) and $u(v_0+v_1)^*x \subseteq L$.

We need some preliminaries. We call a NFA $A$ with $\varepsilon$-transitions {\em simple} if every strongly
connected component (SCC) of $A$ is either trivial or a cycle containing at least one non-$\varepsilon$ transition,
and every bottom SCC contains a final state. Clearly, if $A$ is simple then there
is a finite union $p_1, \ldots, p_n$ of bounded expressions such that $L(A) = p_1 + \cdots + p_n$ (informally,
each $p_i$ corresponds to a path in the acyclic graph obtained by contracting every SCC to a single node).
Conversely, every finite union of bounded expressions is recognized by a simple NFA with $\varepsilon$-transitions.

Since $L$ is regular, there is NFA with $\varepsilon$-transitions $A_L$ such that $L(A_L)= L$.
 W.l.o.g. we can assume that every bottom SCC
of $A_L$ contains some final state. Since $L$ is infinite, $A_L$ contains at least
one nontrivial SCC reachable from the initial state.
Since $L$ is not equal to a finite union
of bounded expressions, $A_L$ contains at least one SCC, say $C$, reachable from the initial state, that
is not a cycle. Moreover, we can assume that from some state $q$ of $C$ there are two paths leading
from $q$ to $q$ that read two different nonempty words $v_0, v_1$ such that
\(v_0v_1\neq v_1v_0\) (otherwise, $C$ can be ``replaced'' by two cycles: one for \(v_0^*\)
and one for \(v_1^*\)). Let $u$ be any word leading to $q$, and $x$ be any word leading from
$q$ to a final state. Clearly, $u(v_0+v_1)^*x \subseteq L$.

We now prove that the emptiness problem of MHPDA modulo $L$ (and so modulo ${\cal C}$) is undecidable by reduction from the
emptiness problem for intersection of CFG the alphabet $\{0,1\}$.
Let $G_1, G_2$ be two CFG. Using closure
of CFL with respect to concatenation and homomorphism, we
can easily construct grammars $G_1', G_2'$ such that $G_i$ accepts $a_1 \ldots a_n \in \{0,1\}^*$
if{}f $G_i'$ accepts the word $u (w_1 \ldots w_n) x$, where $w_j = v_0$ if $a_j =0$, and $w_j = v_1$ if $a_j=1$
for every $j\in [1,n]$. Now, since $L(G_1'), L(G_2') \subseteq u(v_0+v_1)^*x$, we have
$L(G_1') \cap L(G_2') \cap L =
L(G_1') \cap L(G_2') \cap u(v_0+v_1)^*x$, and so  $L(G_1) \cap L(G_2) = \varnothing$ if{}f
$L(G_1') \cap L(G_2') \cap L = \varnothing$. So the emptiness problem of MHPDA modulo $L$ is
undecidable

\noindent
{\bf point 2.}
Since CSL is closed under boolean operations and has a decidable membership problem,
CSL is perfect modulo ${\cal F}$. Any class of regular languages stronger than ${\cal F}$
contains an infinite regular language $L$. We prove that emptiness of CSL modulo $L$ is undecidable by reduction
from the emptiness problem for CSL, which implies that CSL is not perfect modulo $L$.

Since $L$ is infinite, there are words $w_1,w_2,w_3$ such that $w_1 w_2^*w_3 \in L$.
Given a context-sensitive grammar $G$, it is easy to construct a grammar $G'$ satisfying
$L(G') \subseteq w_1 w_2^*w_3$ and such that $L(G)$ is empty if{}f $L(G')$ is empty. First, we
replace every terminal symbol of $G$ by a variable generating $w_2$,
and then we add a new production $S' \rightarrow S_1 S S_3$, where $S$ is the axiom of $G$,
and $S_1, S_3$ are variables generating $w_1, w_3$.
\end{IEEEproof}

\section{Applications to Verification}

In this section, we show MHPDAs are expressive enough to capture several
automata-theoretic models. More surprisingly, we show that MHPDA are an elegant
solution to find optimal complexity results as well.  As an appetizer consider
the non emptiness problem for the intersection of \(k\) context free languages and
a bounded expression \(\pat\). In \cite{ge11}, the authors show that
this problem is in NP, and use it to show that assertion checking of multithreaded
programs communicating through shared memory is in NP as well. To
show that this result is subsumed by ours, proceed as follows. First,
compute in polynomial time  \(1\)-HPDAs \(\set{M_i}_{i\in[1,k]}\) recognizing
the context-free languages. Then, use Thm.~\ref{thm:superemptiness} to compute in
\(O(k\cdot \max_i(|M_i|) \cdot |\pat|^3)\) time a formula \(\Psi\)
such that \(\Psi\) is satisfiable if{}f the intersection of \(k\) CFLs and \(\pat\) is
non empty. Conclude that the problem is in NP.

In the next two sections we prove that the control-state reachability problem for
recursive counter machines (\(\kcm\)) and communicating finite-state machines (\(\cfsm\)) modulo
bounded expressions also reduces to bounded emptiness of MHPDA, and
use this to prove that both problems are NP-complete.

\section{Recursive Counter Machines}\label{sec:models}

Let $k \geq 1$. A \emph{recursive counter machine} ($\kcm$) is a tuple
$(S,\Gamma,\mathcal{C},\trans,s_0)$ where $S$ is a non-empty finite set of
\emph{control states}; $\Gamma$ is a \emph{stack alphabet} with a distinguished
\emph{bottom stack symbol} $\bot$; $\mathcal{C} =\set{c_1,\ldots,c_k}$ is a
finite set of $k$ \emph{counters}; $s_0\in S$ is the \emph{initial} control
state; and $\trans$ is a finite set of \emph{transitions} of the form $(\alpha,
\gamma) \stackrel{\op}{\rightarrow} (\beta, v)$, where $\alpha, \beta\in S$,
$\gamma \in \Gamma$, $v \in \Gamma^*$, and $\op \in\set{\inc_i, \dec_i,
\zerotest_i}_{i\in [1,k]}$ is one of the {\em counter operations} increment,
decrement, or test for zero of $c_i\in \mathcal{C}$ respectively.

A \emph{configuration} $(s,w,v_1,\dots,v_k)\in S\times \Gamma^* \times \nats^k$
consists of a control state $s$, a stack content $w$, and a valuation of the
counters.  The \emph{initial configuration} is $\config_0 = (s_0, \bot,
\pzero)$. Let $t$ be a transition $(\alpha, \gamma) \stackrel{\op}{\rightarrow}
(\beta, v)$.  We say that a configuration $\config'=(s',w',v'_1,\dots,v'_k)$ is
a \emph{flow $t$-successor} of $\config =(s,w,v_1,\dots,v_k)$, denoted by
$\config \, F_t \,\config'$, if $s=\alpha$, $s'=\beta$, $w= \gamma u$ and $w'=
vu$ for some $u \in \Gamma^*$.  We say that $\config'$ is a
\emph{$t$-successor} of $\config$, denoted by $\config \, R_t \, \config'$, if
$\config \, F_t \, \config'$ and either $\op =\inc_i$ and
$(v'_1,\dots,v'_k)=(v_1,\dots,v_k)+\punit_i$, or $\op =\dec_i$ and
$(v'_1,\dots,v'_k)=(v_1,\dots,v_k)-\punit_i$, or $\op =\zerotest_i$ and $v_i=0$
and $(v'_1,\dots,v'_k)=(v_1,\dots,v_k)$.  Given a sequence $\pi\in \trans^*$,
we define $F(\pi)$ recursively as follows: $F(\varepsilon)$ is the identity
relation over configurations, and $F(\pi'\cdot t)=F(\pi')\comp F_t$, where
$\comp$ denotes join of relations.  Given $L\subseteq \trans^*$, we define
$F(L) = \bigcup_{\pi\in L}F(\pi)$.  We define $R(\pi)$ and $R(L)$ analogously.
The set of configurations \emph{reachable through $L$} is
$\post[L]=\set{\config \mid  \config_0 \, R(L) \, \config }$.

The \emph{control reachability problem} for $\kcm$ asks, given a control state $s_f$,
whether $\post[\trans^*]$ contains a configuration with control state $s_f$.
The problem is undecidable even for non-recursive counter machines \cite{min67}.

Given a bounded expression $\pat$ over the alphabet $\trans$ of transitions,
the \emph{control reachability problem modulo }\(\pat\) is the question whether
$\post[\pat]$ contains a configuration with control state $s_f$. We show that
this problem is NP-complete by means of a reduction to the bounded emptiness
problem for sequential MHPDAs.

\subsection{Encoding Counter Machines}\label{sec:enckcm}

Fix a $\kcm$ $(S,\Gamma,\mathcal{C},\trans,s_0)$ with $k$ counters and a
bounded expression \(\pat\) over \(\trans\). We construct  $k+1$ \(1\)-HPDAs such
that $\pi \in \trans^*$ is accepted by all the \(1\)-HPDA if{}f $\post[\pi]$ contains
a configuration with $s_f$ as control state. (Following Prop.~\ref{lem:kcmcapb}, we can then construct
an equivalent $(k+1)$-HPDA if we wish.)

The first PDA $P_0$ checks whether $\config_0 \,  F(\pi) \, \config$ holds for
some configuration $\config$ having $s_f$ as control state. Since
for each transition $t$ of the $\kcm$ the relation $F_t$ exactly
corresponds to the relation induced by the productions of a pushdown automaton,
the construction of $P_0$ is straightforward, and we omit the details.

A word $\pi$ accepted by $P_0$ is consistent with the control flow of the
$\kcm$, but might not be feasible ($\pi$ may zero-test a counter
whose value is not 0, or decrement a counter whose value is 0). Feasibility is checked by PDAs $P_1, \ldots, P_k$.
More precisely, $P_i$ checks that the projection of $\pi$ onto
the operations of $c_i$ is feasible.  We first describe a generic PDA
$P_\surd$ over the alphabet $\{+,-,0\}$, where
``$+$'' encodes increment, ``$-$'' decrement, and ``$0$'' a zero-test,
as a template that can be instantiated to generate $P_1, \ldots, P_k$.

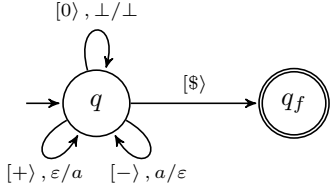
\begin{figure}
	\centering
	\scalebox{1}{
	\begin{tikzpicture}[->,>=stealth',shorten >=1pt,auto,node distance=2.6cm, semithick,initial text=]
  \tikzstyle{every state}=[fill=none,draw=black,text=black]

  \node[initial,state]         (Q)   {$q$};
	\node[accepting,state]    		 (Qf)  [right of=Q] {$q_f$};

	\path[font=\scriptsize]
	(Q) edge node[above] {\(\readh{\rmark}\)} (Qf)
	(Q) edge [loop above] node {\(\readh{0},\bot/\bot\)} (Q)
	(Q) edge [out=210, in=240, loop] node[below] {\(\readh{+},\varepsilon/a\)} (Q)
	(Q) edge [in=300, out=330, loop] node[below] {\(\readh{-},a/\varepsilon\)}     (Q);
\end{tikzpicture}
}
\caption{The \(1\)-HPDA $P_{\surd}$ over alphabet \(\set{+,-,0}\).}
	\label{fig:cm_pda}
\end{figure}

$P_\surd$ is shown in Fig.~\ref{fig:cm_pda}. It
uses its stack as a counter. The stack alphabet is $\set{\bot,a}$.
When $P_{\surd}$ reads a $+$ (a $-$), it pushes an $a$ into (pops an $a$ from) the stack,
and when it reads $0$, it checks that the top element is the end-of-stack marker $\bot$
($\readh{0},\bot/\bot$). Now, $P_i$ is a suitably modified version
which, when reading a letter $t = (\alpha, \gamma) \stackrel{\op}{\rightarrow} (\beta, w)$,
acts according to the operation $\op$: if $\op =\inc_i$ ($\dec_i$, $\zerotest_i$), then
$t$ is treated as $+$ ($-$, $0$). If $\op$ does not
operate on the $i$-the counter, then control ignores $t$.

Applying Thm.~\ref{thm:superemptiness}, we get:

\begin{theorem}{}
\label{thm:reach-modulo-kcm}
Given a $\kcm$ $A=(S,\Gamma,\mathcal{C},\trans,s_0)$ with \(k\) counters, a
control state \(s_f \in S\), and a bounded expression  $\pat$ over $\trans$,
there is a computable formula $\Phi_{A,s_f}$ of existential Presburger arithmetic of size
$O(k\cdot |A|^3 \cdot |\pat|^3)$ such that $\post[\pat]$ contains a
configuration with state $s_f$ if{}f $\Phi_{A,s_f}$ is satisfiable. As a
consequence, the bounded control reachability problem for recursive counter
machines is in NP.
\end{theorem}

NP-hardness holds even for non-recursive counter machines (this result has
been communicated to us by S. Demri, but for completeness a proof can be found
in the Appendix), and therefore the bound of Thm.~\ref{thm:reach-modulo-kcm}
is optimal.

A similar construction can be used to simulate
recursive machines with \(k\)-auxiliary stacks.

\section{Communicating Finite State Machines}\label{sec:cfsm}

Let $k\geq 1$. A communicating finite state machines ($\cfsm$) is a tuple $(S,K,\Sigma,\trans,s_0)$ where
$S$ is a non-empty finite set of \emph{control states}; $K=\set{\channel_1,\dots,\channel_k}$ is a finite set of \emph{unbounded FIFO channels}; $\Sigma$ is a
non-empty finite set of \emph{messages}; $s_0\in S$ is the \emph{initial} control state; and
$\trans$ is a finite set of \emph{transitions}. Each transition $t\in
\trans$ is given by a triple $(\alpha_t,\op_t,\beta_t)$ where $\alpha_t, \beta_t
\in S$ and $\op_t$ is the \emph{channel operation}: either
$\enqueue{\sigma}\colon \channel_i$, which writes  message $\sigma\in \Sigma$ to
channel $\channel_i$ or $\dequeue{\sigma}\colon\channel_j$,
which reads message $\sigma\in \Sigma$ from channel $\channel_j$.
A \emph{configuration} is a tuple $(s,x_1,\ldots,x_k)\in S\times [\Sigma^*]^{k}$
containing a control state and the content of each channel $\channel_i\in K$. The {\em initial
configuration} is $\config_0=(s_0,\varepsilon,\ldots,\varepsilon)$.

Given $t=(\alpha,\op,\beta)\in \trans$, we define the relations $F_t$ and $R_t$ over
configurations as follows: $(s,x_1,\dots,x_k) \mathbin{F_t} (s',x'_1,\dots,x'_k)$
if{}f $\alpha_t = s$ and $\beta_t = s'$, and
$(s,x_1,\dots,x_k) \mathbin{R_t} (s',x'_1,\dots,x'_k)$ if{}f $s=\alpha$,
$s'=\beta$, and for all $i\in[1,k]$ either $x_i=\sigma\cdot x'_i$ and
$\op_t=\dequeue{\sigma}\colon\channel_i$, $x'_i=x_i\cdot \sigma$ and
$\op_t=\enqueue{\sigma}\colon\channel_i$, or $x'_i=x_i$ otherwise.

$F(L), R(L), \post[L]$, the control reachability problem and the control
reachability problem modulo a bounded expression
for $\cfsm$s are defined as for $\kcm$. The reachability problem for $\cfsm$ is undecidable \cite{cfsm1983}.

\subsection{Encoding Communicating Machines}\label{sec:enccfsm}

We proceed as for recursive counter machines. Given a $\cfsm$ with
$k$ channels, we construct a finite automaton $P_0$ and \(k\) $2$-HPDAs $P_1, \ldots, P_k$ such that
$\pi \in \trans^*$ is accepted by all of $P_0, \ldots, P_k$ if{}f $\post[\pi]$
contains a configuration with $s_f$
as control state. Again, $P_0$ checks whether
$\config_0 \,  F(\pi) \, \config$ holds for
some configuration $\config$ having $s_f$ as control state, and
$P_1$ to $P_k$ check feasibility of $\pi$. In the case of $\cfsm$, feasibility
means that the contents of the channels after taking a transition $t$ are the ones
given by $R(t)$.

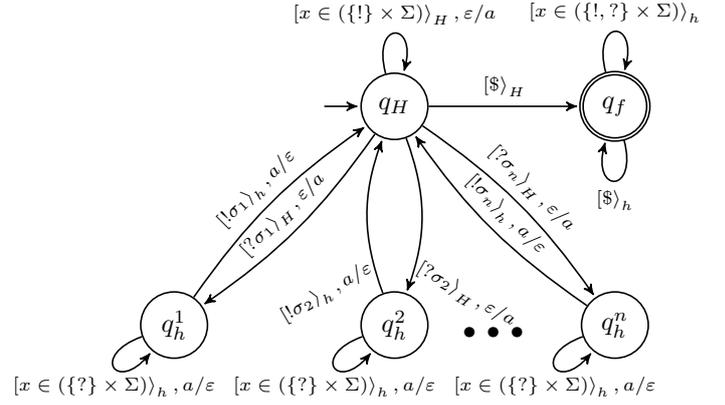
\begin{figure}[t]
	\centering
	\begin{tikzpicture}[->,>=stealth',shorten >=1pt,auto,node distance=2.9cm, semithick,initial text=]
  \tikzstyle{every state}=[fill=none,draw=black,text=black]

  \node[initial,state]         (QH)  {$q_H$};
	\node[accepting,state]    		 (Qf)  [right of=QH] {$q_f$};
	\node[state]         (Qnh) [below of=Qf] {$q_h^{n}$};
	\node[state]         (Q2h) [left of=Qnh] {$q_h^{2}$};
	\node[state]         (Q1h) [left of=Q2h] {$q_h^{1}$};
	\node at (1.3,-3) (O) {$\bullet\,\bullet\,\bullet$};
	\path[font=\scriptsize]
	(QH)  edge [loop above] 	node {\(\readh{x\in (\set{\enqueue{}}\times\Sigma)}_H,\varepsilon/a\)} (QH)
	(Qf)  edge [loop above] 	node {\(\readh{x\in (\set{\enqueue{},\dequeue{}}\times\Sigma)}_h\)} (Qf)
	(Qf)  edge [loop below] 	node {\(\readh{\rmark}_h\)} (Qf)
	(QH)  edge 						 		node {\(\readh{\rmark}_H\)} (Qf)
	(Q1h) edge [out=200, in=230,loop] 	node[below] {\(\readh{x\in (\set{\dequeue{}}\times\Sigma)}_h,a/\varepsilon\)} (Q1h)
	(Q2h) edge [out=200, in=230,loop] 	node[below] {\(\readh{x\in (\set{\dequeue{}}\times\Sigma)}_h,a/\varepsilon\)} (Q2h)
	(Qnh) edge [out=200, in=230,loop] 	node[below] {\(\readh{x\in (\set{\dequeue{}}\times\Sigma)}_h,a/\varepsilon\)} (Qnh)
	(QH) 	edge [bend left=10] node[above,sloped] {\(\readh{\dequeue{\sigma_1}}_H,\varepsilon/a\)} (Q1h)
	(Q1h) edge [bend left=10] node[above,sloped] {\(\readh{\enqueue{\sigma_1}}_h,a/\varepsilon\)} (QH)
	(QH) 	edge [bend left=20,pos=.7] node[rotate=330] {\(\readh{\dequeue{\sigma_2}}_H,\varepsilon/a\)} (Q2h)
	(Q2h) edge [bend left=20,pos=.3] node[rotate=30] {\(\readh{\enqueue{\sigma_2}}_h,a/\varepsilon\)} (QH)
	(QH) 	edge [bend left=10]	node[above,sloped] {\(\readh{\dequeue{\sigma_n}}_H,\varepsilon/a\)} (Qnh)
	(Qnh) edge [bend left=10] node[above,sloped] {\(\readh{\enqueue{\sigma_n}}_h,a/\varepsilon\)} (QH);
\end{tikzpicture}
\caption{The \(2\)-HPDA $P_{\surd}$ where \(\Sigma=\set{\sigma_1,\dots,\sigma_n}\).}
	\label{fig:cfsm_pda}
\end{figure}

$P_0$ is even simpler as for $\kcm$, since there is no
recursion. \footnote{Our results also hold for recursive $\cfsm$, but
since this model is rather artificial we refrain from describing it.}

$P_i$ checks feasibility of $\pi$ with respect to the $i$-th channel.
As in the case of $\kcm$, we define a generic \(2\)-HPDA
$P_{\surd}$, depicted in Fig.~\ref{fig:cfsm_pda}, that checks consistency for a
channel $\channel$.

For convenience the heads of $P_{\surd}$ are named $h$ and $H$.  The stack
alphabet is $\set{\bot,a}$, where $\bot$ is, as above, a special end-of-stack
marker.  $P_{\surd}$ works as follows.
In state $q_H$, head $H$ reads symbols \(\set{\enqueue{\sigma_i}\mid
i\in [1,n]}\) to channel $\channel$ until a symbol $\dequeue{\sigma_i}$ for some
\(i\in[1,n]\) or \(\rmark\) is read .
When \(\dequeue{\sigma_i}\) is read, control jumps to
\(q_h^i\). In \(q_h^{i}\), head $h$ looks
for the first symbol
\(\set{\enqueue{\sigma_i}\mid i\in [1,n]}\). If it is
$\enqueue{\sigma_i}$ (which corresponds to \(\dequeue{\sigma_i}\)) then
control returns to \(q_H\). Intuitively, if a symbol is read from channel \(\channel\) it must have been written previously.
Observe that the stack ensures ensure that \(h\) does not move beyond $H$.
In fact, in every reachable configuration not in state \(q_f\), $P_{\surd}$ maintains the invariant that the number of symbols between $H$ and $h$ coincides with the number
of $a$'s on the stack.  For instance for tapes \(\tuple{t_h,t_H}\) where
\begin{align*}
t_h&=\enqueue{\sigma_1}\#\enqueue{\sigma_2}\dequeue{\sigma_1}\dequeue{\sigma_2}\&\\
t_H&=\enqueue{\sigma_2}\phantom{\#}\enqueue{\sigma_2}\dequeue{\sigma_1}\dequeue{\sigma_2}\& \#
\end{align*}
the stack content is given by $\bot a^3$.
Because of the invariant, head \(H\) will be the first to read \(\rmark\) in
which case the control is updated to \(q_f\).  Hence transitions read anything
until with head \(h\) until it falls down the tape.

We can now apply Thm.~\ref{thm:superemptiness} again. In this case, the members
of our family of MHPDAs have at most 2 heads, i.e., $c=2$.

\begin{theorem}{}
\label{thm:reach-modulo-cfsm}
Given a $\cfsm$ $A=(S,K,\Sigma,\trans,s_0)$ with \(k\) channels, a control
state \(s_f \in S\), and a bounded expression  $\pat$ over $\trans$, there is a
computable formula $\Phi_{A,s_f}$ of existential Presburger arithmetic of size
\(O(k\cdot |A|^3 \cdot |\pat|^6)\)
such that $\post[\pat]$ contains a configuration with state $s_f$ if{}f
$\Phi_{A,s_f}$ is satisfiable.  As a consequence, the bounded control
reachability problem for $\cfsm$ is NP.
\end{theorem}

Again, we can prove that NP-hardness holds for \(\cfsm\),
and therefore that our bound is optimal. The proof is in Appendix.

Finally, let us observe that the above reduction can be extended so as to
handle machines where transitions are either counter operations or channel
operations, i.e.  \((S,\Gamma,\mathcal{C}\cup K,\trans,s_0)\).  The
construction of \(P_0\) is as for \(\kcm\).  Then, for each auxiliary
storage \(S\in \mathcal{C}\cup K\), it suffices to use the adequate MHPDA (for
counter or channel) checking for feasibility of a sequence of operations on
\(S\). Again, we can show an NP upper bound for the bounded control
reachability problem.

\section{Conclusions}
We have introduced verification modulo a class of languages,
which formalizes the common practice, for efficiency reasons, of
checking only a subset of the behaviours of a system.
This leads to the notion of a perfect computational model ${\cal M}$
modulo a class of behaviours ${\cal C}$. We have presented a
perfect model for the class of \emph{bounded expressions}:
multi-head pushdown automata (MHPDA). We have determined the complexity of
the emptiness problem, shown that many popular modelling
formalisms can be easily compiled into MHPDA, and proved that the
compilation leads to verification algorithms of optimal complexity.

There are two interesting open problems. The first one is to search for
more expressive perfect models modulo bounded expressions. The second
is to determine whether our bounds relating the sizes of two MHPDAs accepting
a bounded language and its bounded complement are tight.

\bibliographystyle{IEEEtran}
\bibliography{IEEEabrv,ref}

\newpage

\appendix[Missing Proofs]

\subsection{Proposition~\ref{prop:conexphard}}

\noindent
\rule{\linewidth}{1pt}
The emptiness problem for MHPDAs modulo an arbitrary bounded expression is
in coNEXPTIME. Moreover, the emptiness problem for MHPDAs
and $\pat=(01)^*$ is coNEXPTIME-hard.
\noindent
\rule{\linewidth}{1pt}

\begin{IEEEproof}[Proof of Prop.~\ref{prop:conexphard}]
Membership in coNEXPTIME follows immediately from Thm.~\ref{thm:emptiness}
and the fact that satisfiability of existential Presburger formulas is in NP \cite{GS78}.

For the hardness part, we reduce from $0$-$1$ Succinct Knapsack.

\begin{description}
\item
{\bf Input}: Boolean circuit $\theta$ with $k+n$ variables ($k, n >0$ given in unary).
The circuit represents $2^k$ numbers $a_0,\ldots,a_{2^k-1}$, each with $2^n$ bits in binary,
as follows.
The $i$th bit of the binary representation of $a_j$ is $x\in\set{0,1}$ if
the circuit $\theta$ on input $\mathit{bin}^k(j), \mathit{bin}^n(i)$ evaluates to $x$,
for $i\in [0,2^n-1]$, $j\in [0,2^k-1]$, where $\mathit{bin}^\alpha(\beta)$ is the binary
representation of $\beta$ using $\alpha$ bits.

\item
{\bf Output}: ``Yes'' if there exist $z_1,\ldots, z_{2^k-1} \in\set{0,1}$ such that
$a_0 = \sum_{i=1}^{2^k-1} a_i z_i$; ``No'' otherwise.
\end{description}
Given an instance of the $0$-$1$ Succinct Knapsack problem, we construct in polynomial time a
MHPDA that accepts a string of the form $(01)^*$ if{}f the $0$-$1$ Succinct Knapsack problem answers ``Yes.''

The idea of the proof is to use $d$ heads of a MHPDA and the bounded expression $(01)^*$ to encode
$2^d$ states, and to use the stack to compute up to $2^{2^d}$.
The MHPDA use two heads, one to track $a_0$ and one to track the sum on the r.h.s.
If these heads point to the same location at the end, we accept.
Note that we cannot directly check if two heads are pointing to the same location.
However, we can alternately move the heads to the right (by reading) and check
that they hit the end marker at the same time.

We start with some preliminary constructions.
We use $d$ heads $h_1,\ldots,h_d$ to encode a $d$-bit configuration $b\in\set{0,1}^d$:
to encode $b$, we make sure that head $h_i$ is pointing to bit $b_i$ on the tape.
For $b\in\set{0,1}^d$, we write $b_i$ for the $i$th bit of $b$.
With this representation, we denote by \(\boldsymbol{h}\) $d$-bit binary number given by the symbols
under the heads $h_1,\ldots,h_d$. Also, we denote by \(\denote{\boldsymbol{h}}\) the number that is represented.

Given a constant $c\in \set{0,1}^d$, we can check that the current store
encodes $c$ without destroying the current encoding as follows.  First, observe
that the bounded expression $(01)^*$ ensures that by reading twice from any head, the head
points to the same bit as it was pointing to before the two reads (for a long
enough string). For $i=1,\ldots, d$, read twice with \(h_i\) and remember the
first value, say \(x\), that is read.  Then check that $x=c_i$.
If not, we go to a state signifying that the current configuration is not
storing $c$, otherwise we continue the next iteration of the loop.  At the end
of the loop, we go to a state that signifies that the current encoding is equal
to $c$.  The stack is not touched.

Given heads $h_1,\ldots,h_d$, we can ``reset'' the encoding to a specific
$c\in\set{0,1}^d$ (noted \(\boldsymbol{h}\leftarrow c\)) as follows.  For
$i=1,\ldots,d$, read with \(h_i\) and let \(x\) be the value read.  If
\(x=c_i\), then again read with $h_i$; else do nothing (because after the read
with \(h_i\), it points to \(c_i\)). The stack is not touched.

Given heads $h_1,\ldots, h_d$ and $h'_1,\ldots, h'_d$, we can ``copy'' the
encoding of the $h_i$s to $h'_i$s (\(\boldsymbol{h'}\leftarrow\boldsymbol{h}\)) as follows.
For $i=1,\ldots,d$, we execute the following.  Read twice with \(h_i\) and
remember the first value, say \(x\), that is read.  Now read with \(h'_i\), if
the value read equals \(x\) then read again; else do nothing.  At the end of
updating the $d$ heads we have that \(\boldsymbol{h'}\) equals \(\boldsymbol{h}\).  The stack
is not touched.

Given the binary number $\boldsymbol{h}$, $\boldsymbol{h} \not = 1^d$, we show how to add one
to the number such that the resulting \(\boldsymbol{h}\) encodes
\(\denote{\boldsymbol{h}}+1\). Read with \(h_1\), if the symbol read is \(0\) then we
are done (\(h_1\) points to \(1\)); else (\(h_1\) points to \(0\)) do the
following: read with \(h_2\), if the symbol read is \(0\) (\(h_1\) points to
\(1\)) then we are done.  In general, if $h_i$ points to zero (and all
$h_1,\ldots,h_{i-1}$ point to $1$'s), read with each head $h_1,\ldots, h_i$.
We can similarly subtract one from the number $\boldsymbol{h}$, $\boldsymbol{h} \not = 0^d$,
by replacing zero with one in the above construction.  In both constructions,
the stack is not touched.

Finally, suppose we have heads $h_0,\ldots,h_d$, a head $H$, and $2^d$ bits $c_1c_2\ldots c_{2^d}$
on the stack.
Let $C$ be the number with binary representation $c_{2^d}\ldots c_1$.
We show how the head $H$ can be moved $C$ times to the right, using the heads $h_0,\ldots,h_d$.
Note that $C$ can be as large as $2^{2^d} -1$, so we cannot directly store $C$ using $poly(d)$ heads.
Instead, we use the $d$ heads to count the position in the stack, and perform binary arithmetic
on the number in the stack. We execute the following program.

\begin{tabbing}
\ \ \ \ \=\ \ \ \=\ \ \ \=\ \ \ \=\\
\>\(\denote{\boldsymbol{h}}\leftarrow 2^{d}\)\\
1:\>\textbf{while} \(\denote{\boldsymbol{h}}\neq 0\) and top of stack is 0 \{\\
  \>\> pop;\\
  \>\> \(\denote{\boldsymbol{h}}\leftarrow \denote{\boldsymbol{h}}-1\);\\
  \>\}\\
	\>\textbf{if} \(\denote{\boldsymbol{h}}=0\) \(\{\)\\
  \>\> exit; \verb+/* H has now moved C times */+\\
	\>\(\} \) \textbf{else} \(\{\) \verb+/* top of stack is necessarily 1 */+\\
  \>\> pop 1; push 0; read with \(H\);\\
	\>\> \textbf{while} \(\denote{\boldsymbol{h}}\neq 2^d\) \(\{\)\\
  \>\>\> push 1;\\
  \>\>\> \(\denote{\boldsymbol{h}}\leftarrow \denote{\boldsymbol{h}}+1\);\\
	\>\> \(\}\)\\
  \> \(\}\)\\
	\> \textbf{goto} 1;
\end{tabbing}
Using the constructions above, the program can be implemented by an MHPDA of size polynomial in $d$.
We call this procedure $\mathit{MoveRight}(H)$.

We now show how to evaluate the circuit $\theta$.
W.l.o.g., we assume that $\theta$ is given as $p_1(n+k)$ layers, each layer has $p_2(n+k)$ binary gates, for
polynomials $p_1$ and $p_2$.
We use $k+n + p_1(k+n)p_2(k+n)$ heads.
The inputs are copied into $k+n$ heads.
Then, we evaluate the value of each gate, starting at the lowest layer, and store it into the head
representing that gate.
To evaluate the gate, we look at the values encoded by the heads representing its inputs,
and evaluate the Boolean function for the gate.
The stack is untouched in the evaluation.
Thus, circuit evaluation can be performed by an MHPDA (indeed, a multi-head finite automaton) using polynomially
many (in $k+n$) heads.

Now we come to the main construction.
The MHPDA has the following heads:
\begin{itemize}
\item a head $A_0$ to track $a_0$, a head $\mathit{SUM}$ to track the r.h.s.
\item $k$ heads $K_1,\ldots,K_k$ to track the indices of the numbers $a_0,\ldots,a_{2^k-1}$;
\item $m$ heads $M_1,\ldots,M_m$ to track the $2^m$ bits of each number;
\item $m+1$ heads $H_0,\ldots,H_m$ to implement procedure $\mathit{MoveRight}$ above;
\item additional heads (polynomial in $k+n$) to evaluate circuit $\theta$.
\end{itemize}
Initially, each head points to a $0$, in particular, $\boldsymbol{K} = 0^k$.
The MHPDA works in the following phases.

In the first phase, we initialize $\boldsymbol{M}$ to $1^m$ and then run the following iteratively.
We evaluate $\theta$ on the input $\boldsymbol{K};\boldsymbol{M}$ (by copying \(K_1,\ldots,K_k,M_1,\ldots,M_m\)  on to
the circuit inputs and then evaluating the circuit), and push the evaluated value on to the stack.
If $\boldsymbol{M} = 0^m$ we move to the next phase of the construction.
Otherwise, we subtract $1$ from \(\denote{\boldsymbol{M}}\) and repeat the evaluation.

At the end of the above loop, we have $2^m$ bits, representing the number $a_0$ stored on the stack
(least significant bit on top).
We now invoke $\mathit{MoveRight}(A_0)$, which will move head $A_0$ of $a_0$ times to the right.

Then comes the phase of guessing and summing a subset of
\(\set{a_1,\ldots,a_{2^{k}-1}}\) to compare the resulting value against
\(a_0\). First we set $\denote{\boldsymbol{K}}$ to \(1\). For
$\denote{\boldsymbol{K}} = 1$ to $2^k-1$, do the following loop.  We guess if
$z_{\denote{\boldsymbol{K}}}$ is zero or one, using the finite state of the
automaton.  If $z_{\denote{\boldsymbol{K}}}$ is guessed to be zero, we continue
with the next iteration of the loop.  Otherwise, we initialize $\boldsymbol{M}$
to $1^m$, and iteratively evaluate $\theta$ on $\boldsymbol{K};\boldsymbol{M}$
for each $\boldsymbol{M}$ from $1^m$ to $0^m$, and push each evaluated bit on
the stack.  At the end of the process, we have the $2^m$ bits of
$a_{\denote{\boldsymbol{K}}}$ on the stack, least significant bit first.  We
now invoke $\mathit{MoveRight}(\mathit{SUM})$ to move the head $\mathit{SUM}$
$a_{\denote{\boldsymbol{K}}}$ times to the right.

At the end of the loop, we have that the head $\mathit{SUM}$ has moved
$\sum_{i=1}^{2^k-1} z_i a_i$ times to the right, where the $z_i$'s are the
guesses made by the MHPDA.  We now check if $A_0$ and $\mathit{SUM}$ are
pointing to the same tape cell by moving them alternately and checking that
they read the end marker $\rmark$ immediately one after the other.  If so, we
read with all heads until they fall off the tape and accept.  Otherwise, we
reject.  Note that the computations can be performed by a MHPDA that is
polynomial in the size of the input.

If the answer to the $0$-$1$ Succinct Knapsack instance is ``Yes,'' then there is a
sequence of guesses, and a string in $(01)^*$ that is sufficiently long to
perform all the computations, such that the MHPDA accepts.  However, if the
answer is ``No'' then the language of the automaton is empty.

Thus, given a MHPDA $M$, and the fixed bounded expression $(01)^*$, checking if
$L(M) \cap (01)^*$ is empty is coNEXPTIME-hard.
\end{IEEEproof}

\subsection{Proposition~\ref{prop:letter-bounded-complement}}

\noindent
\rule{\linewidth}{1pt}

Given a $d$-HPDA \(A\) and a letter-bounded expression \(\patb= b_1^* \ldots b_n^*\),
there is an MHPDA \(B\) such that \(L(B)=\patb\setminus L(A)\) and
\(|B|\) is at most doubly exponential in \(|A|\), $|\patb|$.

\noindent
\rule{\linewidth}{1pt}

\begin{IEEEproof}[Proof of Prop.~\ref{prop:letter-bounded-complement}]
The complementation procedure follows these steps:
\begin{itemize}
\item Compute the family $\{\Psi_i(x_1, \ldots, x_n)\}_{i=1}^\alpha$ of
	existential Presburger formulas of Prop.~\ref{prop:emptinessNP}, each of them
	of size $p(|A| \cdot |\patb|)$ for a suitable polynomial $p$. Recall that
	$\alpha = d^{|\patb|d}$.

\item Compute quantifier-free formulas $\Phi_i \equiv \neg \Psi_i$ with
	constants in unary of size $|\Phi_i| \in {\it 2exp}(O(|\Psi_i|))$.
        By Prop.~\ref{prop:emptinessNP} we have
	$b_1^{k_1}\dots b_n^{k_n}\in (\patb\setminus L(A))$ if{}f
	$\bigvee_{i=1}^\alpha \Psi_i(k_1, \ldots, k_n)$ is false if{}f
	$\bigwedge_{i=1}^\alpha \Phi_i(k_1, \ldots, k_n)$ is true.

\item Construct for every formula $\Phi_i$ a MHPDA $B_i$ of size $O(\Phi_i)$ as
	in Prop.~\ref{prop:comp}.
\item Let $B$ be a MHPDA accepting $\bigcap_{i=1}^\alpha L(B_i)$, which exists
	by Prop.~\ref{lem:kcmcapb}.  We have
\begin{align*}
	|B| & \in O(\textstyle{\sum_{i=1}^\alpha} |B_i|) \\
		& \in O(\textstyle{\sum_{i=1}^\alpha} |\Phi_i|) \\
		& \in \textstyle{\sum_{i=1}^\alpha} {\it 2exp}(O(|\Psi_i|)) \\
    & \in d^{nd} \cdot {\it 2exp}(p'(|A| \cdot n)) \\
    &  =  {\it 2exp}(p''(|A| \cdot n))
\end{align*}
\noindent for suitable polynomials $p', p''$.
\end{itemize}
\end{IEEEproof}

\subsection{NP-hardness of control state reachability modulo bounded expressions for Counter Machines}

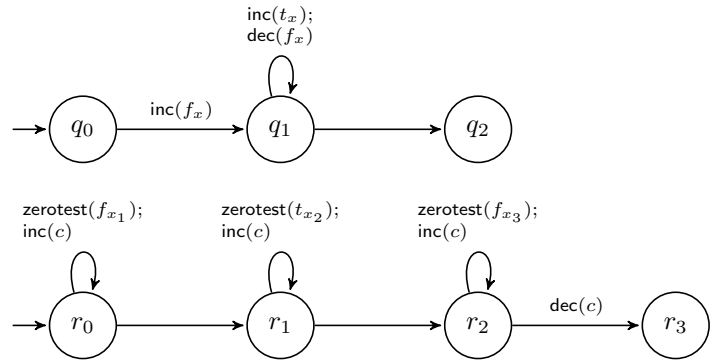
\begin{figure*}[b!]
\centering
\begin{tikzpicture}[->,>=stealth',shorten >=1pt,auto,node distance=2.6cm,semithick, initial text=]
\tikzstyle{every state}=[fill=none,draw=black,text=black]

\node[initial,state] (Q0)		{$q_0$};
\node[state] 	     (Q1) [right of=Q0] {$q_1$};
\node[state] 	     (Q2) [right of=Q1] {$q_2$};
\path[font=\scriptsize]
(Q0) edge node {$\inc(f_x)$} (Q1)
(Q1) edge [loop above] node {$\begin{array}{l}\inc(t_x);\\ \dec(f_x)\end{array}$} (Q1)
     edge node { } (Q2);

\node[initial,state] (C0) [below of=Q0]      { $r_0$ };
\node[state] (C1) [right of=C0] { $r_1$ };
\node[state] (C2) [right of=C1] { $r_2$ };
\node[state] (C3) [right of=C2] { $r_3$ };
\path[font=\scriptsize]
(C0) edge [loop above] node {$\begin{array}{l}\zerotest(f_{x_1});\\ \inc(c)\end{array}$} (C0)
     edge node { } (C1)
(C1) edge [loop above] node {$\begin{array}{l}\zerotest(t_{x_2});\\\inc(c)\end{array}$} (C1)
     edge node { } (C2)
(C2) edge [loop above] node {$\begin{array}{l}\zerotest(f_{x_3});\\\inc(c)\end{array}$} (C2)
     edge node {$\dec(c)$} (C3);
\end{tikzpicture}
\caption{Reduction for CMs. The top gadget shows variable assignment. The bottom gadget
shows the checks for a clause $c \equiv x_1\vee \lnot x_2\vee x_3$. \label{fig:cm-np}}
\end{figure*}

\begin{IEEEproof}
We reduce from 3SAT.
Given a 3SAT formula $c_1 \wedge \ldots \wedge c_m$ over variables $x_1,\ldots,x_n$, we construct
a \(\kcm\) with counters $\set{t_{x_i}, f_{x_i} \mid i\in[1,n]} \cup \set{c_i \mid i\in[1,m]}$.
We use a gadget to assign values to variables and a gadget to check that a clause is satisfied by the current
assignment to variables.
Fig.~\ref{fig:cm-np} shows the gadgets.

For each variable $x$ in the formula, we keep two counters $t_x$ and $f_x$.
The variable gadget (top of Fig.~\ref{fig:cm-np}) ensures that when control reaches
$q_2$, then either $t_x = 1$ and $f_x = 0$ (encoding that $x$ is true) or
$t_x = 0$ and $f_x = 1$ (encoding that $x$ is false), depending on whether the loop is
executed one or zero times, respectively.
Note that the loop can be executed at most once: the second iteration gets stuck decrementing $f_x$.

The clause gadget (bottom of Fig.~\ref{fig:cm-np}) shows how we check that a
clause $c \equiv x_1\vee \lnot x_2 \vee x_3$ is satisfied.
The gadget keeps a ``control'' counter $c$.
The first loop checks that $f_{x_1} = 0$ (i.e., $t_{x_1}=1$, and $x_1$ is set to true) and increments $c$.
The second loop checks that $t_{x_2} = 0$ (i.e., $f_{x_1}=1$, and $x_2$ is set to false) and increments $c$.
The third loop checks that $f_{x_3} = 0$ (i.e., $t_{x_3}=1$, and $x_3$ is set to true) and increments $c$.
Each loop can be executed any number of times.
At the end, the decrement succeeds only when at least one iteration of a loop has executed, which indicates that
$c$ is satisfied.
Note that if $c$ is not satisfied, control cannot reach the last location $r_3$: either one of the tests in
the loops get stuck, or the decrement at the end gets stuck.

For the reduction, we sequentially compose gadgets for all the variables and then all the clauses and ask if
the control state at the end of the last clause can be reached. Clearly, paths of the automaton conform to a
bounded expression.
\end{IEEEproof}

\subsection{NP-hardness of control state reachability modulo bounded expressions for \(\cfsm\)}

\begin{IEEEproof}
We reduce from 3SAT.
Given a 3SAT formula $c_1 \wedge \ldots \wedge c_m$ over variables $x_1,\ldots,x_n$, we construct
a \(\cfsm\) with channels $\set{ x_i, \hat{x}_i \mid i\in[1,n]} \cup \set{c_i \mid i\in [1,m]}$.
There are two messages: $0$ and $1$.
The channel $x_i$ is used to keep a guess for the variable $x_i$.
The channel $\hat{x}_i$ is a ``control channel'' used to ensure only one guess is made.
The control flow graph of the \(\cfsm\) consists of gadgets selecting a value for each variable
and gadgets checking that each clause is satisfied.

The gadget for variables is shown on the top of Fig.~\ref{fig:cfsm-np}.
The gadget first puts a single message $0$ into the control channel $\hat{x}_i$.
It then defines two loops.
The first puts $0$ in the channel $x_i$ (thereby guessing $x_i$ is false)
and flips the control channel by dequeueing the $0$ and enqueueing a $1$.
The second puts $1$ in the channel $x_i$ (thereby guessing that $x_i$ is true)
and flips the control channel as before.
Finally, the edge from $q_2$ to $q_3$ dequeues a $1$ from the control channel.

By the use of the control channel, we note that any execution that reaches
$q_3$ must execute exactly one loop, exactly one time.
When control reaches $q_3$, the control channel $\hat{x}_i$ is empty, and
the channel $x_i$ is either $0$ or $1$.

The gadget for clauses is shown in the bottom of Fig.~\ref{fig:cfsm-np}, for the
particular clause $c \equiv (x_1\vee \lnot x_2 \vee x_3)$ (the general case is immediate).
The gadget for the clause has three loops, one for each literal in the clause.
Each loop checks if the value guessed for the variable matches the literal (i.e.,
the clause is satisfied).
If so, a message is added to the channel $c$.
At the end of the three loops (edge $r_2$ to $r_3$), we check that the control channel
$c$ has at least one message.
By construction, control can reach $r_3$ only when the current guess for the variables
satisfies the clause.
Moreover, the channels $x_i$ are unchanged.

The \(\cfsm\) sequentially composes the variable gadgets and the clause
gadgets, and checks if control can reach the last node of the last clause
gadget.  Clearly, paths of the automaton conform to a bounded expression.
\end{IEEEproof}

\begin{figure*}[t]
\centering
\begin{tikzpicture}[->,>=stealth',shorten >=1pt,auto,node distance=2.6cm,semithick, initial text=]
\tikzstyle{every state}=[fill=none,draw=black,text=black]

\node[initial,state] (Q0)		{$q_0$};
\node[state] 	     (Q1) [right of=Q0] {$q_1$};
\node[state] 	     (Q2) [right of=Q1] {$q_2$};
\node[state] 	     (Q3) [right of=Q2] { $q_3$ };
\path[font=\scriptsize]
(Q0) edge node {$!0:\hat{x}_i$} (Q1)
(Q1) edge [loop above] node {$\begin{array}{l}!0:x_i;\\?0:\hat{x}_i;\\!1:\hat{x}_i\end{array}$} (Q1)
     edge node { } (Q2)
(Q2) edge [loop above] node {$\begin{array}{l}!1:x_i;\\?0:\hat{x}_i;\\!1:\hat{x}_i\end{array}$} (Q2)
     edge node {$?1:\hat{x}_i$ } (Q3);

\node[initial,state] (C0) [below of=Q0]      { $r_0$ };
\node[state] (C1) [right of=C0] { $r_1$ };
\node[state] (C2) [right of=C1] { $r_2$ };
\node[state] (C3) [right of=C2] { $r_3$ };
\path[font=\scriptsize]
(C0) edge [loop above] node {$\begin{array}{l}?1:x_1;\\!1:x_1;\\!1:c\end{array}$} (C0)
     edge node { } (C1)
(C1) edge [loop above] node {$\begin{array}{l}?0:x_2;\\!0:x_2;\\!1:c\end{array}$} (C1)
     edge node { } (C2)
(C2) edge [loop above] node {$\begin{array}{l}?1:x_3;\\!1:x_3;\\!1:c\end{array}$} (C2)
     edge node {$?1:c$} (C3);
\end{tikzpicture}
\caption{Reduction for \(\cfsm\)s. The top gadget shows variable selection. The bottom gadget
shows the checks for a clause $x_1\vee \lnot x_2\vee x_3$. \label{fig:cfsm-np}}
\end{figure*}

\end{document}